\documentclass[11pt]{amsart}

\pdfoutput=1

\usepackage{geometry}                
\usepackage{graphicx}
\usepackage{amssymb}
\usepackage{epstopdf}
\usepackage{ifpdf}
\usepackage{moreverb}
\usepackage{natbib}
\usepackage{pgfplots}
\usepackage{subfigure,lscape}
\usepackage{gensymb}		
\usepackage{tabu}		
\usepackage{colortbl}		

\newcommand\BibTeX{{\rmfamily B\kern-.05em \textsc{i\kern-.025em b}\kern-.08em
T\kern-.1667em\lower.7ex\hbox{E}\kern-.125emX}}
\newcommand{\p}{\mathbf{p}}
\renewcommand{\P}{\mathbf{P}}
\newcommand{\s}{\mathbf{s}}
\renewcommand{\S}{\mathbf{S}}
\newcommand{\B}{\mathbf{B}}
\newcommand{\Y}{\mathbf{Y}}

\newcommand{\btheta}{\boldsymbol{\theta}}

\begin{document}

\title[Tools for predicting rainfall from lightning records]{Tools for predicting rainfall from lightning records: events identification and rain prediction using  a Bayesian hierarchical model}

\author[E. Di Giuseppe]{Edmondo Di Giuseppe}
\author[G. Jona Lasinio]{Giovanna Jona Lasinio}
 \author[M. Pasqui]{Massimiliano Pasqui} 
 \author[S. Esposito]{Stanislao Esposito}
 \thanks{E. Di Giuseppe, National Research Council-Institute of Biometeorology (CNR-Ibimet), Via dei Taurini 19 00185, Rome, Italy E-mail: e.digiuseppe@ibimet.cnr.it}
\thanks{G. Jona Lasinio, Sapienza University of Rome-Department of Statistics (Uniroma1-Dss), P.le A. Moro 5 00185, Rome, Italy E-mail: giovanna.jonalasinio@uniroma1.it}
\thanks{M. Pasqui, National Research Council-Institute of Biometeorology (CNR-Ibimet), Via dei Taurini 19 00185, Rome, Italy E-mail: m.pasqui@ibimet.cnr.it}
\thanks{S. Esposito, Consiglio per la ricerca e la sperimentazione in agricoltura-Research Unit for Climatology and Meteorology Applied to Agriculture (CRA-CMA), Via del Caravita 7/a 00186, Rome, Italy E-mail: stanislao.esposito@entecra.it}

\begin{abstract}
We propose a new statistical protocol for the estimation of precipitation using lightning data. We first identify rainy events using a scan statistics, then we estimate Rainfall Lighting Ratio (RLR) to convert lightning number  into rain volume given the storm intensity. Then we build a hierarchical Bayesian model aiming at the prediction of  15- and 30-minutes cumulated precipitation at unobserved locations and time using information on lightning in the same area. More specifically, we build a Bayesian hierarchical model in which precipitation is modeled as function of lightning count and space time variation is handled using specific  structured (random) effects. The mean component of the model relates precipitation and lightning assuming that the number of lightning recorded on a regular grid depends on the number of lightning occurring in neighboring cells. We analyze several model formulations where  storms propagation speed, spatial dependence and time variation incorporates different descriptions of the phenomena at hand.  The space-time variation is assumed separable.  The study area  is located in Central Italy, where two storms, that differ for duration and intensity, are presented. 
\end{abstract}

\keywords{Scan statistics; conditionally autoregressive models;  MCMC estimation}

\maketitle

\section{Introduction}

Several methods and algorithms have been developed for estimating precipitation from satellite data. These methods can be categorized according to the type of sensors used. Infrared IR-based algorithms use the cloud-top brightness temperature \citep{Arkin:1987}, radar-based  use passive microwave recorded from satellite radiometer \citep{Ferraro:1995,Kummerow:2001,Severino:2005,Li:2008} and a combination of IR and satellite radar are proposed in  \citet{Turk:2000}, \citet{Todd:2001}, \citet{Joyce:2004}, \citet{db-satellite}. However, rainfall fields estimates derived from these sensors have many limitations. More specifically, IR-based algorithms have deficiencies in estimating rainfall fields because they do not allow to "differentiate thin non precipitating cirrus clouds and non raining cold mesoscale convective system (MCS) cloud shields" from precipitating clouds \citep{Morales:2003}. On the other hand, satellite radar estimates over a specific geographical area present a significant overpassing time period  and thus limiting the acquisition frequency \citep{Huo2014}. Combined algorithms reduce such limitations though they are still biased in mountain areas and mid-latitude regions.
The methods used by those algorithms to produce a spatial distribution of precipitation estimates are generally based on linear regression. For instance, \citet{Arkin:1987} obtain rain estimates by means of a linear relationship between cold cloud and rainfall; \citet{Ferraro:1995} adopt a radar-rainfall linear relation on the log scale; \citet{Kummerow:2001} apply a linear regression using an empirical index that links hydro-meteors to IR channels; \citet{Joyce:2004} produce estimates by a time-weighted interpolation on microwave derived precipitation. On the other hand, others algorithms use spatio-temporal models in order to reproduce a rainfall spatial surface: \citet{Severino:2005} use geostatistical models combined with autoregressive components to obtain estimates from rain gages and radar measurements; \citet{Li:2008} perform a calibration of radar measurements using a combination of threshold estimation, bias reduction, regression and geostatistical techniques that account for season, rainfall type, and rainfall amount. \citet{db-satellite} produce estimates of the surface rainfall rates as a combination of the infrared brightness temperature by the GEO-IR satellites and microwave radiometer estimates, by means of Kalman filter technique.

Many attempts to improve satellite precipitation estimate have been done by using lightning data. For example various algorithms based on a combination of IR, lightning and radar have been developed \citep{Goodman:1988,Grecu:2000,Morales:2003, Chronis:2004}. In these works, lightning are used as a variable for cloud patches classification, then brightness temperature (of clouds)-rainfall relation is estimated for each clouds group.
For instance, \citet{Morales:2003} individuate regions with lightning observations that are subsequently used to define the degree of active convection in a cloud. Then, they determine the rainy area fraction on the base of the linear relationship with brightness temperature and use it for assigning rain rates to convective area. \citet{Grecu:2000} use the count of lightning to classify rainy convective systems, then they employ a bivariate linear regression to determine the cluster rain volume. They demonstrate that a combination of lightning and IR brightness temperature could reduce by $15\%$ the error variance of rain volume estimates. On the other hand, \citet{Betz:2008} use lightning data for tracking rainy cells inside convective events and show that this is particularly helpful in anticipating the development of severe weather and strong storm cells. The natural evolution of this work is to associate a rainfall rate to tracked cells.

Another approach involving lightning data for rainfall estimation is  based on the \emph{rainfall lightning ratio} (RLR), i.e. an estimate of the amount of precipitation to be associated to each flash. In this approach lightning data are used  for the identification of rainy convective areas and for calculation of the corresponding lightning-rainfall relation, and then to forecast the precipitation amount in proximal areas and nearest times of single flashes. Among others, the most interesting proposals are those of \citet{Cheze:1997}, \citet{Petersen:1998}, and \citet{Tapia-Smith:1998}. \citet{Cheze:1997} describe the rainfall amount as a function of lightning activity for areas of about  $40000 \,km^2$ and 15-minutes time resolution. \citet{Petersen:1998} link the lightning-rainfall relation to the type of convection, \citet{Tapia-Smith:1998} build a model for rainfall estimation on an area of $25\,km^2$ and 5-minutes resolution. Finally, \citet{Tuomi2005} use a "CellSearch" clustering procedure that group flashes to identify thunderstorms and their lifetime characteristics  after a careful algorithm tuning. Notice that none of these  papers deals with the uncertainty associated to rainfall-fields  estimation.
Our model builds upon the RLR  approach. In particular we start from the work of \citet{Tapia-Smith:1998} and following three steps we propose a Bayesian spatio-temporal hierarchical model to obtain rainfall-fields estimates. The three steps are: (i) identification of rainfall \emph{convective events} through a scan  statistic similar to the one propose in \citet{Tuomi2005}; (ii) estimation of rainfall lightning ratio distinguishing among different classes of storms size. In other words from identified convective events we estimate the rainfall lightning ratio taking into account the magnitude of the convective event in terms of total lightning counts; and (iii) rainfall estimation from lightning records through a Bayesian spatio-temporal hierarchical model including an evolution of the basic Tapia-Smith-Dixon model into the mean term of the stochastic model.  

We propose a space-time hierarchical Bayesian model in which precipitation are modeled, on the log scale, as function of lightning counts  and space time variation is handled using specific random effect. We envision a model in which the rain at cell \textit{p} of a regular grid, at time \textit{t} is described by a latent random process adjusting for several specific issues of the available data. Our final goal is to build a predictor for the underlying spatial surface of the latent variable. Furthermore, the adoption of a stochastic approach let us measure  the error associated to predictions. Notice that in what follows we define \emph{convective event} an area where one or more convective systems are present for a given amount of time.
The study area  is located in Central Italy and the chosen case study events are the storms of August 5, 2004 and  May 9, 2006, for both we analyze two time scale of data aggregation: 15- and 30-minutes.
The model is implemented using JAGS \citep{jags}. \\
The database and two case study are presented in Section \ref{sec:Data_Case_study} where we describe both step (i) and (ii), i.e.  the scan statistic procedure used to identify convective events and the RLR estimation. In Section \ref{sec:TheModel} we introduce the model implementation and the estimation of both posterior parameters distribution and posterior predictive distribution (step (iii)).  Section \ref{sec:Results} reports estimation outputs and  an evaluation of model predictions and a comparison with satellite-derived estimates. Finally, discussion and concluding remarks are in Section \ref{sec:Conclusions}.

\section{Available data and case-studies}
\label{sec:Data_Case_study}

\subsection{Data}
\label{subsec:Data}

The study area is identified by the geographical coordinates $41\degree$-$45\degree$ LatN, $9.5\degree$-$14\degree$ LonE. The database is obtained by merging three databases covering the time span  March-September 2003-2006 and reporting  lightning instantaneous records, satellite hourly precipitation fields on a $10x10$ km regular grid and the weather stations sub-hourly precipitation records.

\begin{itemize}

\item Lightning data report the location and date of all registered cases of Cloud to Ground (CG) lightning within the study area. We use \citet{db-fulmini} database acquired by Consorzio Lamma-Regione Toscana and Cnr-Ibimet \citep{db-stazioni}. The CESI-Sirf lightning detection network is formed by 16 sensors positioned on the Italian peninsula and 8 sensors located in Austria, France and Switzerland. The lightning signal is an electromagnetic field detected by at least 3 sensors and, successively, cleaned from the background noise. The optimal distance between sensors is about $200\,km$ although, they have a good performance up to $500\,km$. The sensors are of type IMPACT with broadband electromagnetic antennas and GPS synchronization (Global Atmospherics Technology Inc.).

\item Global satellite precipitation data are distributed by the project GSMaP by the Earth Observation Research Center, Japan Aerospace Exploration Agency (JAXA). The project GSMaP is sponsored by JST-CREST and promoted by the JAXA Precipitation Measuring Mission (PMM) Science Team \citep{Okamoto:2005,db-satellite2,db-satellite3}.The \texttt{$GSMaP_MVK+$} dataset has a grid resolution of $0.1$ degree Lat/Lon and a temporal resolution of 1 hour. The estimates of the surface rainfall rates are obtained as a combination of the infrared brightness temperature by the GEO-IR satellites and microwave radiometer estimates, by means of Kalman filter technique \citep{db-satellite}.

\item Point precipitation data are composed of sub-hourly observations time series coming from $316$ weather stations located within the study area \citep{db-stazioni}. However, we removed all rain gages that presented more the $10 \%$ of missing values on the 15 (30)-minutes time scale. Then, the number of selected rain gages reduce to $171$ for the event of August 5, 2004 of which $159$ returns  15-minutes cumulate rainfall recordings and $12$  30-minutes cumulate rainfall recordings. For   the event of May 9, 2006 we have $181$ rain gages of which $179$ returning 15-minutes recordings and $2$ collecting 30-minutes cumulate rainfall. 
\end{itemize}

Notice that rain gage, lightning and satellite recordings  time coding differs. In fact, the gages rainfall recorded at time \textit{t} indicates the rain accumulated during the time interval $(t-1, t]$ whilst the satellite rainfall record as well as the number of flashes at time \textit{t} represents the rain accumulated and the number of flashes recorded during the time interval $[t, t+1)$, respectively. Hence, the three databases require spatial and temporal alignment to be used jointly.

In what follows we superimpose the grid defined by satellite data, i.e. a regular grid with $10\times 10 \,km$ cells to the study area. This grid will be our reference for the spatial alignment of  data. Thus, we refer to the number of lightning counted in each grid cell and we aggregate recording of rain gages belonging to the same grid cell by taking their median value.

Finally, the space-time support of two case study is: $ 100\, (104) $ cells and $ 36\, (18) $ time units for August 5, 2004 15 (30)-minutes event and $ 111\,(112) $ cells and $ 68\,(34) $ time units for May 9, 2006 15 (30)-minutes event. The final database is composed of lightning and rainfall records registered in $ 3600\,(1872) $ and $ 7548\,(3808) $ space-time units for August 5, 2004 15 (30)-minutes and for May 9, 2006 15 (30)-minutes, respectively.

\subsection{Identification of convective events and estimation of the RLR: Step (i) and Step (ii)}
\label{subsec:ScanStatistics}
To properly model lightning-rain relation we need to identify single  convective events. This complex operation is usually done by analyzing the atmospheric circulation by means of satellite images or other large scale instruments. For instance, \citet{Adler:1988}, and \citet{Anagnostou:1997} have used the cloud top temperature obtained by means of infrared satellite data (IR). In practice, they separated convective from stratiform regions in the cloud system by means of IR brightness temperature. 
On the other hand, \citet{Petersen:1998} determined convective regions analyzing the correlation between total lightning flash rates and convective rainfall.  Here, we propose to simplify the operation by using  the time sequence of lightning records to this aim and, particularly, the use of a scan statistic procedure proposed by \citet{dbscan:1996}. The procedure includes three steps:
\begin{itemize}
\item[1)]  we identify a daily significant lightning aggregation using the marginal distribution of hourly lightning counts over the gridded study region; 
\item[2)] in order to capture events taking place around the boundary between two subsequent days our time-window extends from $6$ pm of the previous day to $6$ am of the next, allowing for a 6 hours overlap between adjacent days. For instance, referring to the 9th of May 2006 as a day with a great lightning activity, we extend the analysis from  $6$ pm of May 8, 2006 to $6$ am of May 10, 2006;   
\item[3)]  we apply the scan statistic procedure proposed by \citet{dbscan:1996}  in order to identify the beginning and the end of each convective event generated inside the chosen time-window. The procedure proposed by \citet{dbscan:1996} first locates a sphere of  \emph{radius} $r$  on one record at location $(x,y)$ at time $t$ and includes all records within the chosen radius, if inside the sphere a minimum number of lightning $n_l$ is recorded,   the record is included in the event.  
\end{itemize}

Notice that the scan statistic depends on the distance among points with coordinates in our study given by the spatial ones and time. Geographic coordinates are expressed in UTM (Universal Transverse Mercator) system, zone 32 and time is an integer,  the three dimensional coordinate system has components of very  different order of magnitude then we choose to  scale them (centering with respect to the mean value and dividing by standard deviation).
To choose the value of scan statistic radius we compute it using $r=0.1,0.2,0.3,0.4$ always with $n_l=10$ following empirical consideration on the lightning physics.  The smallest $r$ values return a too fragmented description of the event, while $r=0.4$ aggregates clearly separated events. Hence we fix $r=0.3$.
 The \emph{dbscan} algorithm is included in the R package \emph{fpc} \citep{Henning:2010} and it is a simplified version of the original algorithm which is based on K-dist criterion \citep{dbscan:1996}. For the purpose of clustering different convective events which are eventually occurring during the same day, we also experimented the \textit{spatgraphs} R-library \citep{spatgraphs}, which compute a general adjacency of a given point pattern. We set a geometric adjacency determining a spatial clustering of lightning by means of a connection radius. However, the clustering obtained by applying this alternative method presents some limitations for several tested convective events, the main being the clustering of lightning falling in the same area but at very different time points. 
 
The scan statistic procedure identifies $ 767 $ events with at least $ 50 $ CG-lightning and the largest event counts $ 33364 $ lightning. We define four categories of convective events based on the total number of lightning: \textit{Small}, \textit{Medium}, \textit{Large} and \textit{Very Large} events. These categories are described in Table \ref{tab:part2_events_dimension}, where the number of cases in each class is shown. 
%


\begin{table}[!htbp]
\caption{Classes of dimensionality of convective events defined on the basis of lightning number.}
\label{tab:part2_events_dimension}
\begin{center}
\begin{tabular}{lrr}
\hline
Dimension&\multicolumn{1}{c}{Number of Lightning}&\multicolumn{1}{c}{Number of cases}\\
\hline
Small & $\leq 170$ &$403$\tabularnewline
Medium & $(170,900]$ &$270$\tabularnewline
Large & $(900,8000]$ &$84$\tabularnewline
Very Large &$> 8000$& $10$\tabularnewline
\hline
\end{tabular}
\end{center}
\end{table}



Once we identify single convective events and their magnitude we can  improve on the estimation of the RLR as proposed in \cite{Tapia-Smith:1998}. The RLR determines the mass of rain associated to each flash, this mass is expressed in $kg \; m^{-2}$.  In general, the RLR depends very much on the thunderstorm type and region. Quantitative estimations of RLR  have been proposed in several studies \citep[see][and references therein]{Soula:2001}. The simplest estimator $\hat{Z}$ of RLR is an average of the  RLR estimated for each convective event, i.e. $\hat{Z} = \dfrac{1}{M_{n}} \sum _{e=1}^{M_{n}} \tilde{Z}_{e}$, where $M_{n}$ is the total number of convective events identified and $\tilde{Z_{e}}$ is defined as:

\begin{equation}\label{eq:RLR_single}
\tilde{{Z}_{e}}= \frac{\sum_{h=1}^{T_{h}}\sum_{\p\in Q} r^{SAT}_{h}(\p)}{\sum _{t=1}^{T_{h}}\sum _{\p \in Q} L(t,\p)}
\end{equation}
where $L(t,\p)$ is the number of lightning recorded at time \emph{t} and cell \emph{$\p$}, $Q$ is the entire area covered by the event and  $r^{SAT}_{h}(\p)$ is the satellite precipitation accumulated at cell $\p$ during the 1-hour $h$ time interval. Notice that the time interval is 1-hour since this is the time scale of satellite database, with $h=1, \cdots, T_h$ being $T_h$ the duration of the event in hours. 

Our proposal is to estimate a different value of the RLR for each class of  events size: Small, Medium and Large as  in Table \ref{tab:part2_events_dimension} where \emph{Large} and \emph{Very Large} are merged together. 
We propose a $\hat{Z}$ estimate that is based on satellite data rather than rain gages data, as the latter are not enough to cover the entire study area. In particular the RLR $\tilde{Z}_e$ is calculated as in Eq. \ref{eq:RLR_single} dividing the total volume of precipitation registered in the area  by the total number of lightning registered in the same area, and rain gages sparsity do not let an accurate computation of the numerator. 
Weather stations data can be used to build empirical correction factors accounting for the tendency of satellites  to underestimate rainfall rates and a consequent excess of zero-grid cells in the study area. Here we define two multiplicative correction factors: $f_1=E(r^{STAT}_h-r^{SAT}_h)$ is the average difference between rainfall recorded at weather stations paired with the corresponding satellite record; $f_2= \frac{N^{SAT}}{N^{STAT}}\frac{n^{STAT}_{zero}}{n^{SAT}_{zero}}$ where $N^{SAT}, N^{STAT}$ are the total number of observations from satellite and weather stations and $n^{STAT}_{zero}, n^{SAT}_{zero}$ are the corresponding zero counts. Hence we obtain for convective event size: 
\begin{equation}\label{eq:RLR_modified}
\hat{Z}^d = \dfrac{1}{M_{n}^{d}} \sum _{e=1}^{M_{n}^{d}} f_1\cdot f_2\cdot \tilde{Z}^d_{e}.
\end{equation}

RLR estimated on our data  are reported in Table \ref{tab:part2_RLR} where RLR values are expressed in $10^{3}$ $m^{3}$ per CG-flash. 

\begin{table}[!h]
\caption{Corrected RLR estimates for 3 classes of dimensionality: Small, Medium and Large convective events.}
\label{tab:part2_RLR} 
\begin{center}
\begin{small}
\begin{tabular}{lrrrr}
\hline
RLR ($10^{3}$ $m^{3}$)&\multicolumn{1}{c}{Small events}&\multicolumn{1}{c}{Medium events}&\multicolumn{1}{c}{Large events}&\multicolumn{1}{c}{Entire}\tabularnewline
\hline
Median&$ 0.1$&$  0.6$&$  8.6$&$  0.2$\tabularnewline
Mean&$ 0.4$&$  2.6$& $ \mathbf{24.1}$&$  4.0$\tabularnewline
St.Dev.&$1.1$&$8.7$&$32.3$&$14.5$\tabularnewline
\hline
\end{tabular}
\end{small}
\end{center}
\end{table}

\subsection{Two case study}
\label{subsec:StudyEvents}
\begin{figure}

\subfigure[]{
	\includegraphics[width=0.6\textwidth]{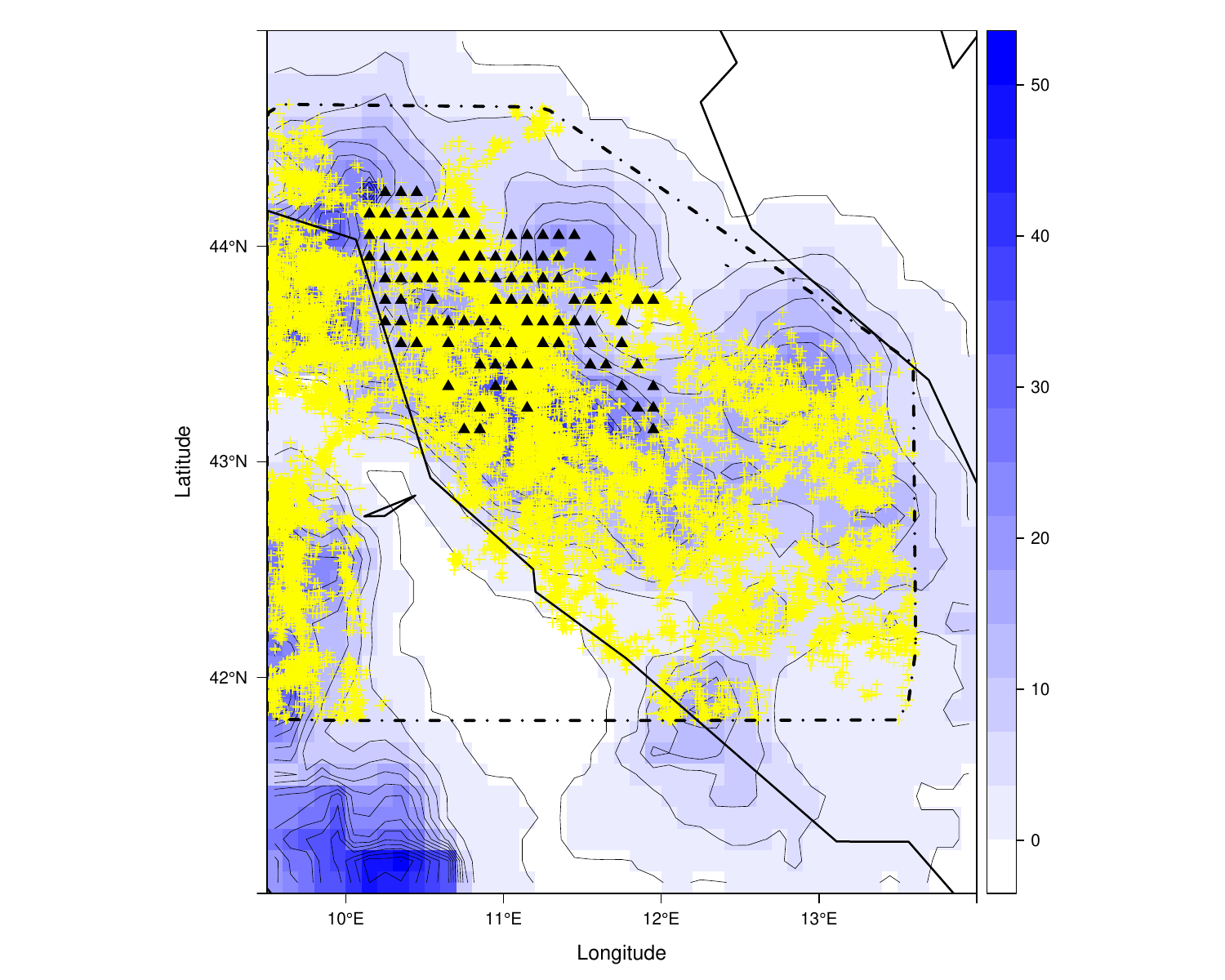}
	}
\subfigure[]{
	\includegraphics[width=0.6\textwidth]{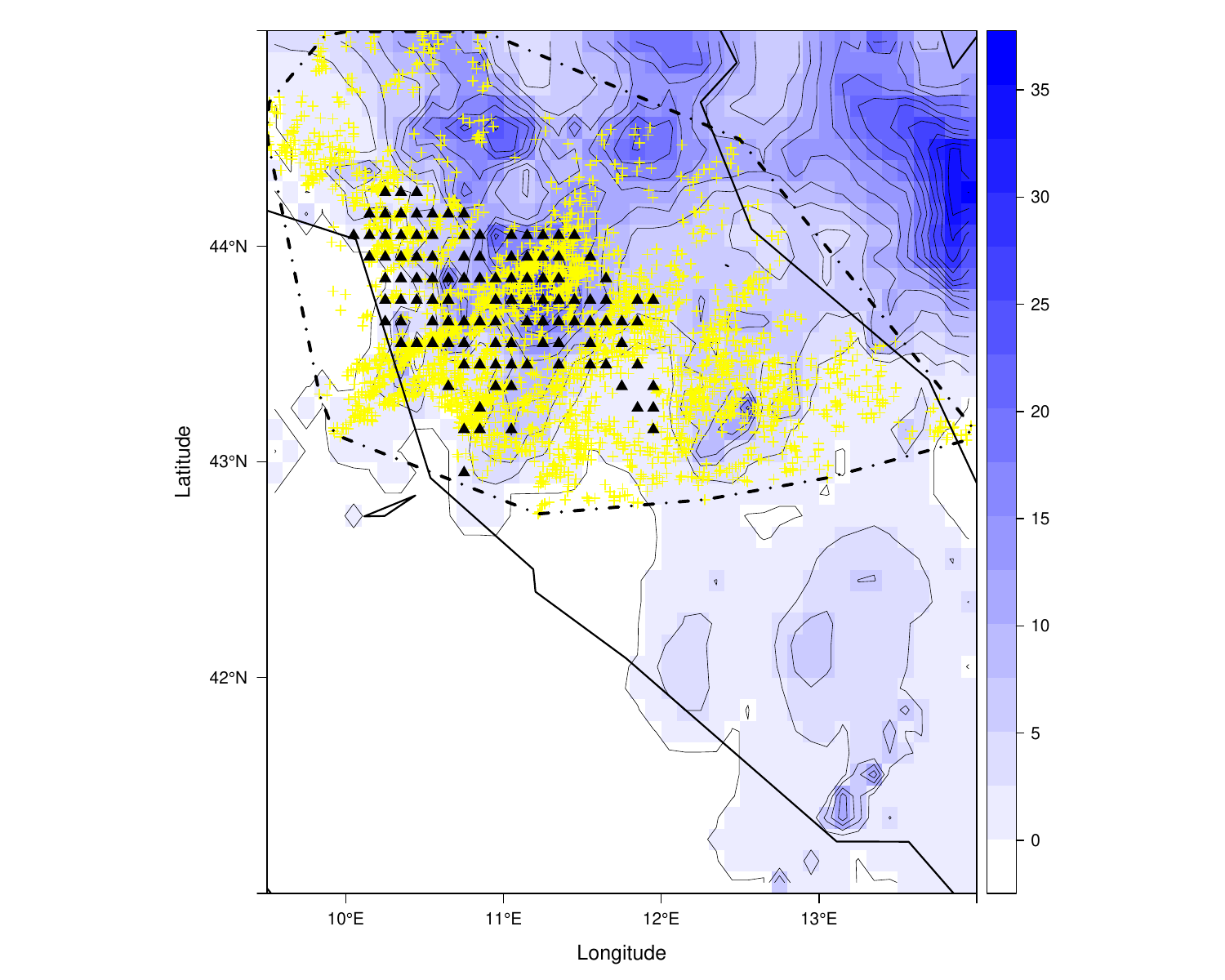}
	}
\caption{The area covered by convective events of August 5, 2004 (a) and May 9, 2006 (b):  polygon (dotted-dashed line) delimiting the event, lightning (light/grey), satellite precipitation (continuous blue scale) and rain gages observations (triangles).}
\label{fig:part2_event_complete}       
\end{figure}

In what follows we model  two convective events occurred during the storms of August 5, 2004 and May 9, 2006. The first convective event started at \emph{11.24}am and ended at \emph{7.56} pm on the 5th of August 2004 and the second event  started at \emph{00.01}am and ended at \emph{4.14} pm on the 9th of May 2006. These events are identified as described in Section \ref{subsec:ScanStatistics}. A map of the two events is reported in Figure \ref{fig:part2_event_complete} where a polygon (black dot-dashed line) delimits the area covered by each convective event and lightning spatial propagation (light yellow), satellite precipitation (continuous blue scale) and observations from rain gages (black triangles) are shown.

From a phenomenological point of view, the two events are quite different since the event of August 5, 2004 is a combination of two type of systems merging together: one is a propagating system generated over Ligurian Sea and entering towards inland, the other is a typical afternoon thermo-convective system generated by orography, whereas the event of May 9, 2006 belongs to a propagating system that enters from Ligurian Sea and stops at Appenini's Mountain chain.  The size of the covered areas, $93147\,km^2$ for August 5, 2004 and  $54131\, km^2$ for May 9, 2006  differs substantially.  August 5, 2004 event's centroid is located at $11.4$ degrees Longitude East and $43.1$ degrees Latitude North and the May 9, 2006 centroid is at $11.4$ degrees Longitude East and $43.8$ degrees Latitude North. The two events differ from each other also because of the total number of lightning generated since the first registered, $ 18140 $ CG-flashes during $ 9 $ hours of duration in August  and  $ 3163 $ CG-flashes during $ 16 $ hours of duration in May. According to the definition given in Table \ref{tab:part2_events_dimension}, the convective event of August 5, 2004 can be considered as a \textit{Very Large event}  whereas that of May 9, 2006 belongs to the category of \textit{Large event}.

We analyze two different cumulate rain: (i) the 15-minutes cumulate  that strictly follows the temporal evolution of lightning within a storm, this measure is affected by a large variability that reflects on the model estimates; (ii) the 30-minutes cumulate that, on the other hand,  removes a large part of the data variability,  but might miss some relevant features of the storm evolution. A comparison of basic summaries of positive precipitation shown in Table \ref{tab:part2_summary_nozero} confirms the different nature of the two events: the mean volume of precipitation calculated as the fraction of total rainfall estimated from satellite data in the area during 15 (30)-minutes on number of cells is $1.386 \,(2.194)\,mm$ for August 5, 2004 and $ 1.299 \,(1.971)\,mm $ for May 9, 2006; the maximum value registered for the 15 (30)-minutes cumulate is $ 34.6\,(58.8)\, mm $ for August 5, 2004 and $ 15\,(17.6)\, mm $ for May 9, 2006. \\
The difference between the two events are stronger  in terms  of extremes of the rainfall distributions (3rd quartile and higher). Indeed the August 5, 2004 event has larger 4th quartile than the May 9, 2006 storm, suggesting larger rain rates. Furthermore, the lightning activity during the peak on the 15-minutes time scale differs substantially since $ 1236 $ flashes have been registered in August 5, 2004 whilst only $ 136 $ flashes have been registered in  May 9, 2006. Notice that a strong difference in the physical structure as well as in the spatial pattern of the two study events allows us to test the model performance under very different circumstances.


\begin{table}
\caption{Summaries of non zero precipitation records of 15- and 30-minutes cumulated rainfall for the two case study.}
\label{tab:part2_summary_nozero}
\begin{center}
\begin{small}
\begin{tabular}{lrrrrrr}
\hline
Event&\multicolumn{1}{c}{Min.}&\multicolumn{1}{c}{1st Qu}.&\multicolumn{1}{c}{Median}&\multicolumn{1}{c}{Mean}&\multicolumn{1}{c}{3rd Qu.}&\multicolumn{1}{c}{Max.}\tabularnewline
\hline
August 5, 2004
15-min&$0.2$&$0.2$&$0.6$&$1.386$&$1.2$&$34.6$\tabularnewline
30-min&$0.2$&$0.4$&$0.8$&$2.194$&$2$&$58.8$\tabularnewline
\hline
May 9, 2006
15-min&$0.2$&$0.2$&$0.6$&$1.299$&$1.6$&$15$\tabularnewline
30-min&$0.2$&$0.2$&$1$&$1.971$&$2.6$&$17.6$\tabularnewline
\hline
\end{tabular}
\end{small}
\end{center}
\end{table}

\begin{figure}
\subfigure[]{
\includegraphics[width=0.4\textwidth]{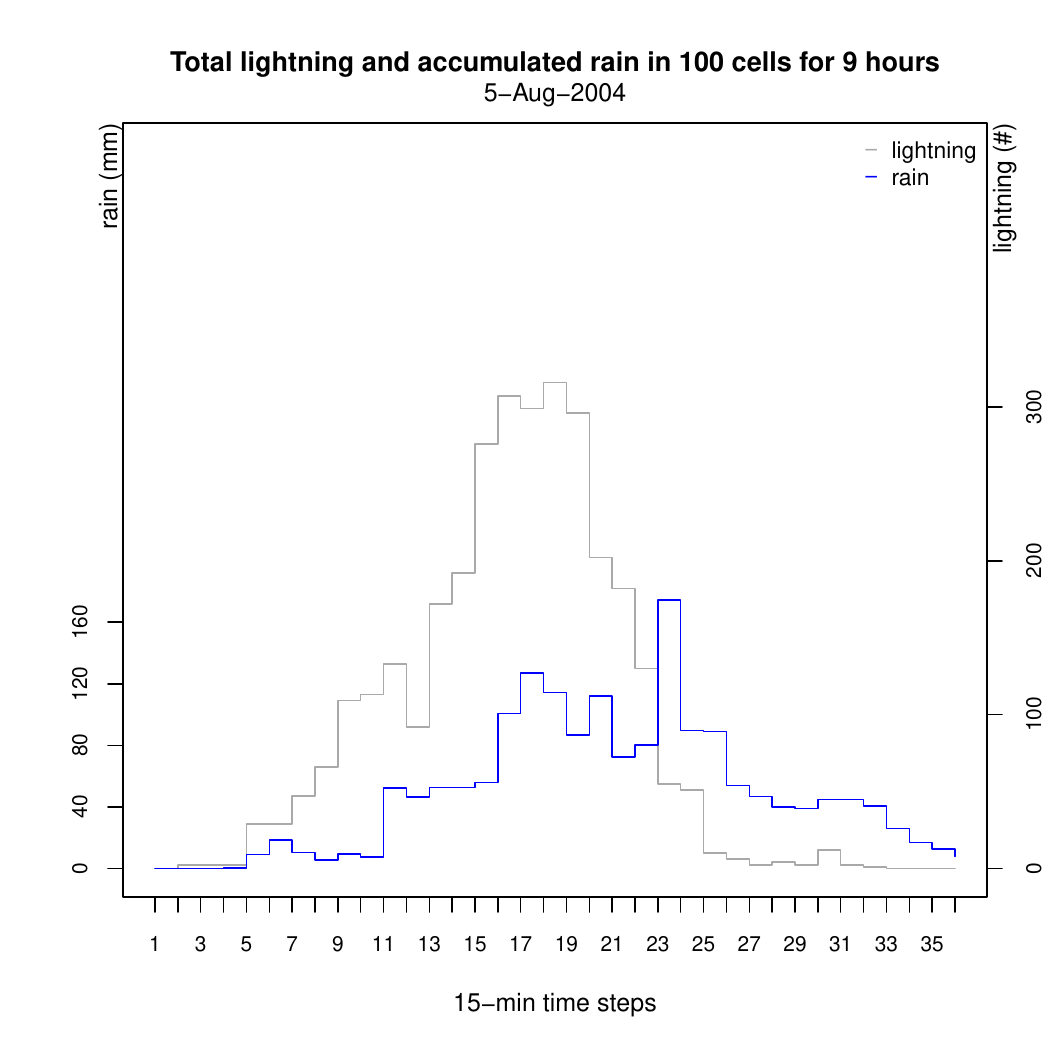}
}
\subfigure[]{
\includegraphics[width=0.4\textwidth]{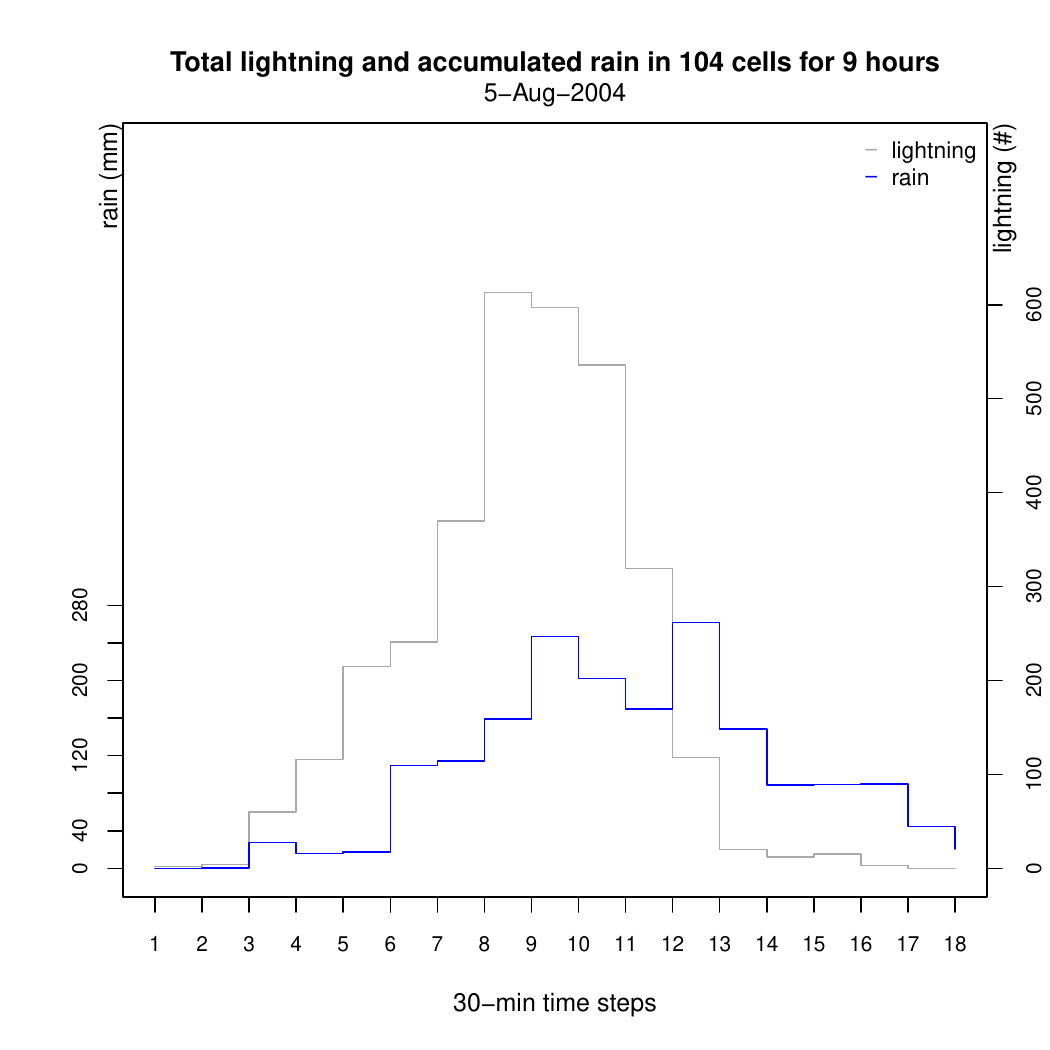}
}\\
\subfigure[]{
\includegraphics[width=0.4\textwidth]{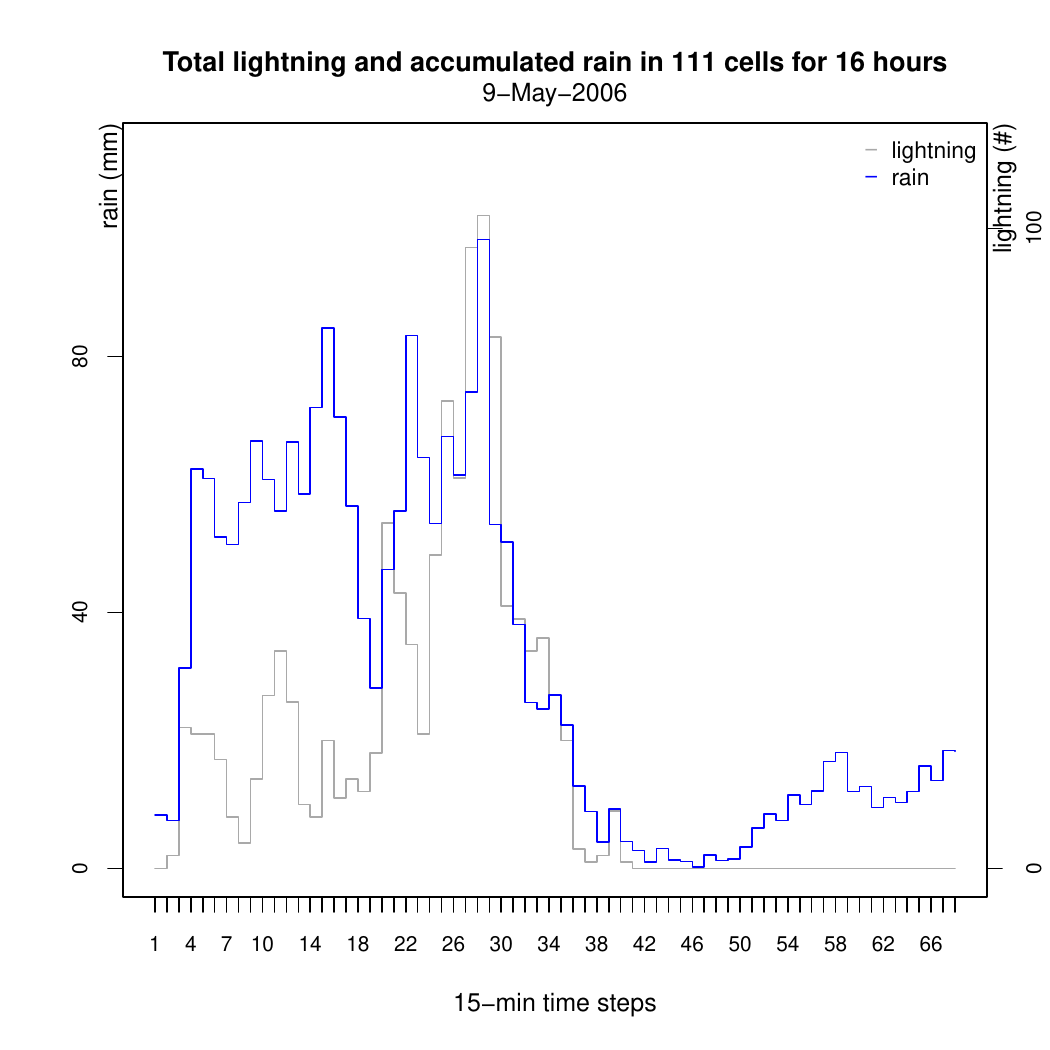}
}
\subfigure[]{
\includegraphics[width=0.4\textwidth]{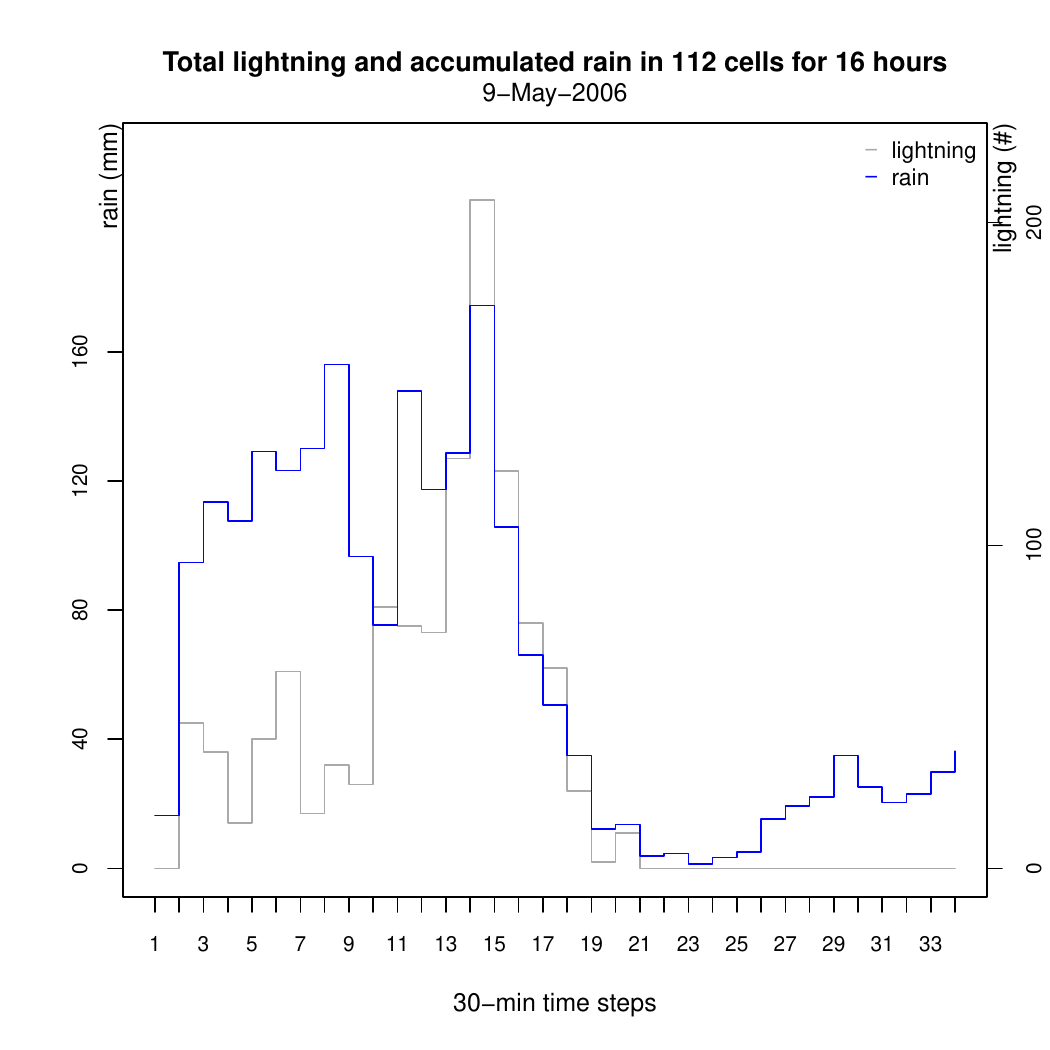}
}
\caption{Time series of rain (blue) and lightning number (light gray) at two time scales 15- and 30-minutes for 5 Aug, 2004 a-b)  and for  9 May, 2006 c-d).}
\label{fig:RainLig_TimeSeriesPlots}       
\end{figure}

In Figure \ref{fig:RainLig_TimeSeriesPlots}, the time dynamic of lightning and rainfall rates at the 15- and 30-minutes time scale is described using the total number of lightings recorded at cell where a rain gage is located and the total rainfall recorded by the latter. The two events differ in their dynamic in the first part of the event time span, after the peak of the storm they behave in a very similar way.

Remark that rainfall data are affected by several problems, on one hand a large number of zeros  is recorded in both time scales, on the other hand rain gages precision (between $ 0.1$ and $0.2$ mm) implies an almost discrete measurement of accumulated rain as shown in Table \ref{tab:part2_cells_summary}. This issue must be taken into account in the modeling of rain gages recording.

\begin{table}
\caption{Frequency distribution of rainfall values observed on 5th of August 2004 and 9th of May 2006 convective events at 15- and 30-minutes time aggregation.}
\label{tab:part2_cells_summary}
\begin{center}
\begin{small}
\begin{tabular}{lrrrrrrr}			
\hline
Rain classes (mm)&\multicolumn{1}{c}{$[0,0.2)$}&\multicolumn{1}{c}{$[0.2,0.4)$}&\multicolumn{1}{c}{$[0.4,0.6)$}&\multicolumn{1}{c}{$[0.6,0.8)$}&\multicolumn{1}{c}{$[0.8,1)$}&\multicolumn{1}{c}{$[1,5)$}&\multicolumn{1}{c}{$\ge 5$}\\
\hline
5Aug2004-15min&$2384$&$380$&$202$&$136$&$102$&$323$&$73$\tabularnewline
\%&$  66.2$&$ 10.6$&$  5.6$&$  3.8$&$  2.8$&$  9$&$ 2$\tabularnewline
5Aug2004-30min&$1079$&$155$&$98$&$89$&$54$&$312$&$85$\tabularnewline
\%&$  57.6$&$  8.3$&$ 5.2$&$ 4.8$&$ 2.9$&$ 16.7$&$ 4.5$\tabularnewline
\hline
9May2006-15min&$5949$&$451$&$202$&$134$&$125$&$625$&$62$\tabularnewline
\%&$  78.8$&$  6$&$  2.7$&$  1.8$&$  1.7$&$  8.3$&$ 0.8$\tabularnewline
9May2006-30min&$2746$&$254$&$120$&$69$&$54$&$446$&$119$\tabularnewline
\%&$  72.1$&$  6.7$&$  3.2$&$ 1.8$&$ 1.4$&$ 11.7$&$  3.1$\tabularnewline
\hline
\end{tabular}
\end{small}
\end{center}
\end{table}


\section{The model: : Step (iii)}
\label{sec:TheModel}

In view of the considerations at the end of previous section we assume that the observed cumulate rainfall at time $t$ and location $\mathbf{p}=(p_x,p_y)$ is a partially discrete stochastic process  $H(t,\mathbf{p})$ such that there exists  a continuous latent process $X(t,\p)$ linked to $H(t,\p)$ as follows \citep{Sahu:2005}:

\begin{equation}\label{eq:obsdiscrete}
H(t,\p)=\left\{\begin{array}{ccc}
\lambda_1^*& \mbox{if} &X(t,\p)<c_1\\
\lambda_2^*& \mbox{if} &c_1\le X(t,\p)<c_2\\
 \vdots &&\\
 \lambda_k^*& \mbox{if} &c_{k-1} \le X(t,\p)<c_k\\
 X(t,\p) &\mbox{if} & X(t,\p) \ge c_k
 \end{array}\right.
 \end{equation}

We assume that there exist $k=5$ values  $\lambda^*_i=\exp(\lambda_i)-1, i=0,\ldots,4$  described in Table \ref{tab:part2_discretization}, that occurs with positive probability. We choose the $\lambda^*_k$ to be the mid point of the discretization intervals. In the sequel we are going to model the latent process on the log scale  ($Y(t,\p)=\log(X(t,\p)+1)$\footnote{We add 1 to preserve zero values.} ) rather than the observed process $H(t,\p)$. This transformation is chosen mostly to smooth the impact of strong rainfall intensities \citep{Zawadzki:2005} and to allow a more sensible adoption of Gaussian representation. This simple method proposed by \citet{Sahu:2005} and \citet{jona:2007} let us to obtain a good performance of the model in predicting zero rainfall events. Other methods can be adopted to treat zero inflated rainfall distributions, such as in \citet{Berrocal:2008}, \citet{Fuentes:2008} or \citet{Schmidt:2009}. \citet{Berrocal:2008} specify a spatial model that includes two spatial Gaussian processes driving precipitation occurrence and accumulation, respectively. A spatial-temporal
model for rain gages and reflectivity radar data is developed in \citet{Fuentes:2008}, where a latent process corresponding to the true rain amount drives the probability of precipitation occurrence and the rainfall accumulation. On the other hand, \citet{Schmidt:2009} treat observations from rain gages as generated by a latent process which realizations are a mixture of a Bernoulli distribution that specifies the probability of having positive precipitation and a probability density function for the rainfall accumulation, typically an exponential, a gamma or a log-normal distribution. Notice that we are working on a finer time scale than the above cited authors that mostly analyze daily and even weekly data. 

\begin{table}
\caption{Discretization values for the latent rainfall field on the log scale $Y(t,\p)$.} 
\label{tab:part2_discretization}
\begin{center}
\begin{tabular}{lr}
\hline
Rain Classes (mm)&\multicolumn{1}{c}{Discretization values}\tabularnewline
\hline
$[0,0.2)$&$\lambda_0=\log(0.1+1)$\\
$[0.2,0.4)$&$\lambda_1=\log(0.3+1)$\\
$[0.4,0.6)$&$\lambda_2=\log(0.5+1)$\\
$[0.6,0.8)$&$\lambda_3=\log(0.7+1)$\\
$[0.8,1)$&$\lambda_4=l\log(0.9+1)$\\
\hline
\end{tabular}
\end{center}
\end{table}


 We assume that 

\begin{equation}\label{eq:likelihood} 
Y(t,\p)|\btheta, \mathbf{W}\sim N\Big(\mu(t,\p)+w(t,\p), \tau^2 \Big)
\end{equation}
where $\mu(t,\p)$ is a  function describing the relation between rainfall and lightning, $w(t,\p)$ is a zero mean separable space-time process, $\tau^2$ is the process variance, and $\btheta$ is the set of all model parameters.  
In the following paragraphs we are going to describe  each model component.
%

\subsection{The mean}
\label{sec:part2_fixed_effect}
To build the model for the process mean we start from the  Tapia-Smith-Dixon deterministic model \citep{Tapia-Smith:1998}. 
The Tapia-Smith-Dixon model is basically a spatio-temporal prediction of rainfall rate based on CG-lightning patterns and RLR estimate, and is formalized as follows:
\begin{equation}
X(t,\mathbf{s})=C \sum _{i=1}^{N_{t}} \, Z \, f(t,T_{i}) \, g(\mathbf{s},\mathbf{S}_{i})
\label{eq:part2_tapia_smith}
\end{equation}
where $X(t,\s)$ is again the observed rainfall rate (mm/h) at time  \textit{t} and spatial location \textbf{\textit{s}},
$C $ is the units conversion factor\footnote{Notice that RLR  is expressed in $kg \; m^{-2}$. whereas the precipitation recorded by rain gages or estimated from satellite data is expressed in $mm \; m^{-3}$. Consequently, we need to transform a mass into a volume applying a conversion factor 
$C=10^{6}A^{-1}$ where \textit{A} is the event area in square kilometers.},  $N_{t}$ is the number of flashes until time $t+\Delta t/2$, $Z$ is the  rainfall-lightning ratio introduced above, $T_{i}$ is the  time at which the  \emph{i}th flash is recorded,  $\mathbf{S}_{i}$ is the location  of the \emph{i}th flash,
$f(t,T_{i})$ specifies the rainfall flux at time \textit{t} determined by a lightning flash at time $T_{i}$ and 
$g(\s,\S_{i})$ specifies the rainfall flux at location $\s$ when a lightning flash is observed at location $\S_{i}$. 
Both  temporal $f(t,T_i)$ and spatial $g(\s,\S_i)$ functions are taken to be \emph{uniform} over the interval $(t-5 , t+5 )$ minute  and within a circle of $5 \,km$ radius around location $\s$, respectively. These assumptions according to the authors are simple and represent a first step to build on. We have already built on the approach of Tapia-Smith-Dixon by introducing estimates of the RLR that depend on the size of convective events (see Eq. \eqref{eq:RLR_modified}). 
We introduce here the results obtained by applying the modified version of Tapia-Smith-Dixon model \citep{Tapia-Smith:1998} through Eq. \eqref{eq:RLR_modified} on the entire set of events identified by scan statistics. Thus, for each event, we estimate precipitation fields at every cell of the spatial domain for the entire duration of the event. This extension of the work is a valid test to evaluate the efficiency of proposed RLR estimator.
The evaluation of this estimation procedure is done using the information from rain gages. More specifically, for the $767$ convective events identified by scan statistic we select those cells where at least one rain gage is settled in, then we compare  rainfall fields from  GSMap satellite data and values computed by implementing mean values of RLR estimated by Eq. \eqref{eq:RLR_modified} into  Eq. \eqref{eq:part2_tapia_smith} with the corresponding rain gage observations. The Root Mean Square Errors (RMSE) values  are reported in Table \ref{tab:RMSE_reconstruction} where "occurrences" are the number of cells with a valid record coming from rain gages.

\bigskip
\begin{table}
	\begin{center}
		\begin{tabular}{lrrr}
			\hline
			\multicolumn{1}{l}{\textbf{{\small RMSE(mm})}}&\multicolumn{1}{c}{{\small RLR=0.4}}&\multicolumn{1}{c}{{\small RLR=2.5}}&\multicolumn{1}{c}{{\small RLR=24.1}}\tabularnewline
			\hline
			Estimated&$ 15.3$&$  24.9$&$  \mathbf{15.9}$\tabularnewline
			GsMap&$ 13.0$&$  22.8$&$  \mathbf{20.7}$\tabularnewline
			\hline
			Occurences&$535$&$1248$&$2780$\tabularnewline
			\hline
		\end{tabular}
	\end{center}
	\caption{Root Mean Square Error of satellite precipitation (GsMap) and estimated fields compared to rain gage observations for 3 category of Small, Medium and Large events. Occurrences are the number of cells with a valid record.\label{tab:RMSE_reconstruction}} 
\end{table}

\smallskip

\begin{table}
	\begin{center}
		\begin{tabular}{lll}
			\hline
			\multicolumn{1}{l}{\textbf{RLR Large events}}&\multicolumn{1}{c}{}&\multicolumn{1}{c}{}\tabularnewline
			\hline
			&POD&POFD\tabularnewline
			Estimated&81.3&15.8\tabularnewline
			GsMap&76.1&15\tabularnewline
			\hline
		\end{tabular}
	\end{center}
	\caption{Comparison of Probability of Detection (POD) and Probability of False Detection (POFD or FAR) indexes between Estimated and GsMap values for Large events.\label{tab:part2_POD_FAR_reconstruction}} 
\end{table}

The analysis of RMSE values reveals that the adoption of our $\hat{Z}(d)$ estimate let us to obtain an important improvement with respect to GsMap satellite data for Large convective events. Furthermore, the probability values of hitting rainy cases, i.e. with rainfall greater than $0.2 \,mm$ and Probability of False Detection (POFD)\footnote{Details on the meanings of POD and POFD are given further in Table \ref{tab:part2_POD_theory}.} confirm this procedure improves the performance of satellite data. \\

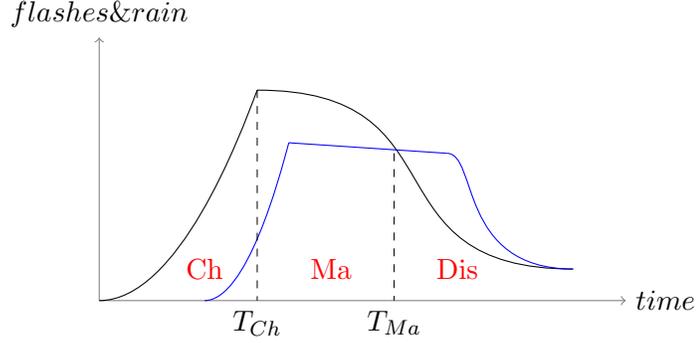
\begin{figure}
\begin{center}
\begin{tikzpicture}[scale=1.4]
    \draw [thin, gray, ->] (0,0) -- (0,2.5)      
        node [above, black] {$flashes\&rain$};              

    \draw [thin, gray, ->] (0,0) -- (5,0)      
        node [right, black] {$time$};              

    \draw  (0,0)  parabola (1.5,2);	    
    \draw  (1.5,2) .. controls (3.5,2) and (2.5,0.3) .. (4.5,0.3);

    \draw [color=blue,thin] (1,0)  parabola (1.8,1.5);	
    \draw [color=blue,thin] (1.8,1.5) -- (3.3,1.4);
    \draw [color=blue,thin] (3.3,1.4) .. controls (3.6,1.4) and (3.4,0.3) .. (4.5,0.3);   

   \draw [dashed, ultra thin] (1.5,2) -- (1.5,0);
   \node [below] at (1.5,0) {$T_{Ch}$};               
   \draw [dashed, ultra thin] (2.8,1.4) -- (2.8,0);
   \node [below] at (2.8,0) {$T_{Ma}$};               
   \node [red] at (1,0.3) {Ch};
   \node [red] at (2.2,0.3) {Ma};
   \node [red] at (3.4,0.3) {Dis};
\end{tikzpicture} 
\end{center}
\caption{Temporal evolution of lightning (black) and associated rain (blue) within a convective event. The 3 stages of evolution are also indicated: \textit{Charging phase} (Ch), \textit{Mature state} (Ma) and \textit{Dissipating phase} (Dis).}
\label{fig:part2_temporal_function}
\end{figure}

Now recall that we are modeling  $Y(t,\p)=\log(X(t,\p)+1)$ the log-latent rainfall process that will be analyzed using two different cumulated amounts: 15- and 30-minutes. We consider time series of length $T$ on a regular grid of $N=n_1\times n_2$ cells. We denote by $L(t,\p)$ the number of lightning at time $t$ and location $\p$. We start by adapting Eq. \eqref{eq:part2_tapia_smith} to estimate rainfall amount: 

\begin{equation} \label{eq:part2_model}
\hat{X}^{LIG}(t,\p) =  (C^* A_{p}^{-1}) \, \hat{Z}^d \, \sum _{i=1}^{T}  \sum_{s \in N_{p}} \; L_{i,\p} \, f(t,T_{i}) \, g(\p,\P_{s})
\hspace{1.5cm}	i=1,2,\cdots, T   
\end{equation}
where $\hat{X}^{LIG}(t,\p)$ is the (latent) rainfall amount predicted at cell $\p$ and time \textit{t}; $\P_{s}$ is the observed cell-location; $T_{i}$ is the observed time;
$L_{i,\p}$ is the number of lightning cumulated at the end of a given time  interval \textit{i} at cell $\p$;
$N_{p}$ is the set of cells belonging to the neighborhood of cell  $\p$; $\hat{Z}^d$ is the estimated Rainfall Lightning Ratio distinguishing the event dimension \textit{d}=Small, Medium, Large as defined in Table \ref{tab:part2_RLR};   $C^*A_{p}^{-1}$ is a conversion factor from $10^{6} kg \; m^{-2}$ (mass) to $mm \; m^{-3}$ (volume) with  \textbf{$A_{p}$} being the area of any cell in square meters and $C^*$ a dimensionality adjusting factor;  $f(t,T_{i})$ is a time weights function;
$g(\p,\P_{s})$ is a spatial weights function.
Time and spatial weight functions are built on the basis of both time evolution and spatial propagation of lightning. In particular, the life of lightning pattern inside a single rainfall convective system is composed of 3 stages: \textit{Charging phase} (Ch), \textit{Mature state} (Ma) and \textit{Dissipating phase} (Dis) (see Figure. \ref{fig:part2_temporal_function}). Here we analyze a convective event that is composed of several single systems, to include the phenomenon evolution in the modeling process, we mimic over the entire event the behavior of the single one. This idea follows the events description given in Figure \ref{fig:RainLig_TimeSeriesPlots}.

%

More precisely we partition the  event duration  into $[t_{0},T_{Ch})$, $[T_{Ch},T_{Ma})$ and $[T_{Ma},T]$.   

\begin{figure}
\begin{center}
\begin{tikzpicture}[scale=.8]

       \draw[step=5mm, black] (0,0) grid (5,5); 
       \draw[black,thick] (0,0) rectangle (5,5);
       \fill[blue] (2.05,2.05) rectangle (2.45,2.45); 

       \fill[green] (1.55,2.05) rectangle (1.95,2.45); 
       \fill[green] (2.55,2.05) rectangle (2.95,2.45); 
       \fill[green] (2.05,2.55) rectangle (2.45,2.95); 
       \fill[green] (2.05,1.95) rectangle (2.45,1.55); 
       \fill[red] (1.55,2.55) rectangle (1.95,2.95); 
       \fill[red] (2.55,2.95) rectangle (2.95,2.55); 
       \fill[red] (2.95,1.95) rectangle (2.55,1.55); 
       \fill[red] (1.55,1.95) rectangle (1.95,1.55); 
\end{tikzpicture}
\end{center}
\caption{Neighborhood of a generic cell \textit{p} (blue):  The first order structure is composed by the green colored  pixels  whilst  the second order structure is obtained by adding the red pixels.}
\label{fig:part2_spatial_function}
\end{figure}
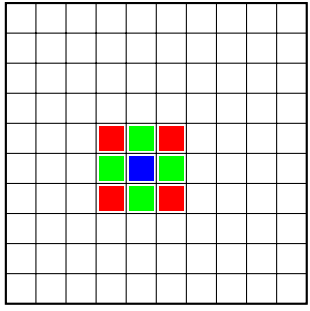

Then, the time weight function is built from  Figure \ref{fig:part2_temporal_function} such that:

\begin{equation}
\label{eq:part2_temporal1}
  f(t,T_{i}) = \;
  \begin{cases}
   f_{Ch}(t) \qquad  & \text{if } \quad t\; \& \;T_{i} \in Ch \\
   f_{Ma,Dis}(t) \qquad  & \text{if } \quad t\; \& \;T_{i} \in Ma \bigcup Dis 
  \end{cases}
\end{equation}
\bigskip
where $T_{Ch}$ indicates the end of the \textit{Charging phase} and \textit{T} is the duration of the entire event. In practice, we assume that the \textit{Mature} and \textit{Dissipating} phases are equivalent in terms of the lightning temporal evolution, than we have  two stages $[t_{0},T_{Ch})$ and $[T_{Ch},T)$. Notice that if the predicting time \textit{t} and observed time $T_{i}$ are in different phases of event's lifetime, that is both events $\bigg\{\{t \in Ch\}\; \cap \; \{T_{i} \in Ma\cup Dis\}\bigg\}$ and $\bigg\{\{t \in Ma\cup Dis\}\; \cap \; \{T_{i} \in Ch\}\bigg\}$, then they receive zero weight, i.e. $f_{Ch;Ma,Dis}(t)= 0$. This assumption is basically made to avoid a strong correlation between model predictions that are distant in time from each other. However, the assumption becomes too strong when dealing with instants of observation and prediction that are both close to the $T_{Ch}$ time. To find a solution for this  drawback, we analyze the performance of  two model modifications in which  the cell's activity is treated without linking it to a specific event's phase. These modifications are discussed below.\\
To build  the space weights function  we assume that the number of lightning in cell $\p$ depends on the number of lightning occurring in neighboring cells. Then we define a neighborhood structure, $N(p)$ to handle such dependency.
In the following we choose $N(p)$ to be a second order neighborhood of cell $p$ (see Figure \ref{fig:part2_spatial_function}), then 
\begin{equation}\label{eq:part2_spatial_function}
  g(\p,\P_{s}) =
  \begin{cases}
   \omega_{i,s}	& \text{if } \P_{s} \in N(p) \\
   0 		& \text{if } \P_{s} \notin N(p) 
  \end{cases}
\end{equation}

\smallskip

\noindent where $\omega_{i,s}=1/8 \, L_{i,\P_{s}}$, $L_{i,\P_{s}}$ is the number of lightning recorded in the observed cell $P_{s}$ at time \textit{i} and the proportional factor $8$ represents the maximum number of neighbors of cell $\p$ for the chosen neighborhood structure. 
In this formulation, the spatial weight between neighboring cells is:
\begin{equation}\label{eq:part2_spatial_function_weights}
\omega_{i,p}=\dfrac{L_{i,p}} {L_{p}} + \frac{1}{8}* \dfrac{L_{i,N_{p}}}{L_{p}} 
\end{equation}
\noindent where $L_{i,p}$ is the number of lightning at predicting cell \textit{p} and time \textit{i}, $L_{i,N_{p}}$ is the summation of lightning over the neighborhood cells of \textit{p} at each time \textit{i} (i.e. $\sum_{P_{s} \in N_{p}} L_{i,P_{s}}$) and
 $
L_{p}=\sum_{i=1}^{T} \big( L_{i,p} + L_{i,N_{p}} \big) 
$
is the total number of lightning hitting cell \textit{p} for the entire duration of the event.

Given the physics behind the propagation of convective events, we are going to  consider several modifications of the mean term of the model.

\paragraph{M1:}
This is the simplest model among those proposed. It includes a very simple form of the temporal weights, that mimic the curves combined in Figure \ref{fig:part2_temporal_function}. To simplify the notation, let us write $C=(C^*A_{p}^{-1})* \hat{Z}^d$ where $C$ is  estimated outside the model as described in section \ref{subsec:ScanStatistics}. Thus, the fixed component of the mean is:

\begin{equation}
\label{eq:mu_ActiveTimes_NoPhysics}
\begin{split}
\mu(t,\p) & = \log \Bigg(  C \sum _{i=1}^{T}  \; L_{i ,p}* \\
	&\bigg( \exp \bigg\{-\big|\frac{t-T_{i}}{\Delta t} \big| \bigg\} I_{[0,T_{Ch}]}(t) 
		 + \exp \bigg\{-\big|\frac{t-T_{i}}{\Delta t} \big|^2 \bigg\} I_{[T_{Ch},T]}(t) 
		  \bigg) \\ 
		& + \; C \sum _{i=1}^{T} \;\omega_{i,p} \; +1 \Bigg)
\end{split}		
\end{equation}

where $I_{[0,T_{Ch}]}(t)$ and $I_{[T_{Ch},T]}(t)$ are indicator functions 
and $\omega_{i,\p}$ is the spatial weight specified in Eq. \eqref{eq:part2_spatial_function_weights}.

\paragraph{M2:}
\label{par:M2}
Here we include the velocity of propagation of the convective event into the temporal weights. We assume this velocity to be a fixed constant ($16.1 \;m/s$) as suggested in \cite{Levizzani:2010} for convective events spanned within $1000\, km$ and up to $20$ hours of duration. Consequently, the time weight functions of Eq. \ref{eq:part2_temporal1} becomes:
\begin{eqnarray} 
\label{eq:part2_temporal2_V}
f_{Ch}(t) & = &  \exp \bigg\{- \frac{(a+bV)} {A^{1/2}_{p}}  \, |\frac{t-T_{i}}{\Delta t} \big| \bigg\}	\qquad \qquad   t_{0} < t\; \& \; T_{i} < T_{Ch} \nonumber \\ \vspace{4cm}
\bigskip
\label{eq:part2_temporal3_V}
f_{Ma,Dis}(t) & = &  \exp \bigg\{- \frac {(a+bV)}  {A_{p}^{1/2}} |\frac{t-T_{i}}{\Delta t} \big|^2 \bigg\} \qquad  T_{Ch} \leq t \;\& \; T_{i}  < T  \nonumber
\end{eqnarray}
where $V$ is the velocity of propagation and $A_{p}$ is the area of a single cell. The mean of model $\mathbf{M}_1$ is:
\[
\label{eq:part2_mu_V}
\begin{split}
\mu(t,\p) &= \log \Bigg(  C \sum _{i=1}^{T}  \; L_{i ,p}* \\
	&\bigg(\exp \bigg\{- \frac {(a+bV)}  {A_{p}^{1/2}} |\frac{t-T_{i}}{\Delta t} \big| \bigg\} I_{[0,T_{Ch}]}(t)
		+ \exp \bigg\{- \frac {(a+bV)}  {A_{p}^{1/2}} |\frac{t-T_{i}}{\Delta t} \big|^2 \bigg\} I_{[T_{Ch},T]}(t)\bigg) \\
		 & + C \sum _{i=1}^{T} \omega_{i,\p} \, +1 \Bigg)
\end{split}
\]
 where $I_{[0,T_{Ch}]}(t)$ and $I_{[T_{Ch},T]}(t)$ are the indicator functions such as in M1.

\paragraph{M3:}
\label{par:V1V2}
in this model the propagation velocity of the convective event is estimated outside the model but on the specific event. In particular, we compute  $V_1$ (Charging phase), $V_2$ (Mature-Dissipating phase) and $V$ (speed of the entire event). The  time weight functions in Eq. \ref{eq:part2_temporal1} become:
\begin{eqnarray} 
\label{eq:part2_temporal2_V1}
f_{Ch}(t) & = &  \exp \bigg\{- \frac{(b_1\,V_1)} {V}  \; \big|\frac{t-T_{i}}{\Delta t} \big| \bigg\}	\qquad   t_{0} < t\; \& \; T_{i} < T_{Ch} \nonumber \\ \vspace{4cm}
\bigskip
\label{eq:part2_temporal3_V2}
f_{Ma,Dis}(t) & = &  \exp \bigg\{- \frac {(b_2\,V_2)} {V} \; \big|\frac{t-T_{i}}{\Delta t} \big|^2 \bigg\} \qquad  T_{Ch} \leq t \;\& \; T_{i}  < T  \nonumber
\end{eqnarray}
$V_1, V_2$ are computed dividing the distance covered by the convective event from  the  starting location (centroid of lightning hit locations in the interval $[t_0,\pm 60sec]$) to peak time location and from peak time to the ending location by the length of each time interval. $V$ is calculated as the ratio between the entire distance covered by the event and its  duration. Then, the latent process mean $\mu(t,\p)$ is:

\begin{equation} 
\label{eq:part2_mu_V1_V2}
\begin{split}
\mu(t,\p) &= \log \Bigg(  C \sum _{i=1}^{T}  \; L_{i ,p}*\\
	&\bigg( \exp \bigg\{- \frac {(b_1\,V_1)} {V} \; \big|\frac{t-T_{i}}{\Delta t} \big| \bigg\} I_{[0,T_{Ch}]}(t) 
		 + \exp \bigg\{- \frac{(b_2\,V_2)} {V}  \; \big|\frac{t-T_{i}}{\Delta t} \big|^2 \bigg\} I_{[T_{Ch},T]}(t) 
		  \bigg) \nonumber \\ 
		 &+ \; C \sum _{i=1}^{T} \;\omega_{i,\p} \; +1 \Bigg)
\end{split}
\end{equation}

 

\paragraph{M4:}
this modification of the model is based on single cell activity rather than on the three lifetime stages of lightning propagation. More specifically, for each cell, firstly we apply a 3-terms moving average on the time series of lightning then we define active only those times where the number of lightning is greater than 5 and we define a cell to be active for 30-minutes following the last selected time to filter out local convection from moving convective systems. These threshold values are similar to those identified by \citet{Tuomi2005} over Finland.  Thus, the formalization of mean component becomes:
\[
\label{eq:mu_ActiveTimes_simple}
\mu(t,\p) = \log \Bigg(  C \sum _{i=1}^{T}  \; L_{i ,p}* 
	\bigg( \exp \bigg\{- \big|\frac{t-T_{i}}{\Delta t} \big| \bigg\} \bigg) \, I_{[t',t'']}(t,\p)  
		 +  C \sum _{i=1}^{T} \;\omega_{i,p} \, I_{[t',t'']}(t,\p) \; +1 \Bigg)
\] 
where $I_{[t',t'']}(t,\p)$ is the indicator function which is $1$ if either the predicted time \textit{t} or the observed time $T_{i}$ are in the phase of cell's activity identified by the time interval $[t',t'']$ and $0$ otherwise. We adopt both simple and quadratic exponential decay, so defining a further model \textit{M5}:
\[
\label{eq:mu_ActiveTimes_quadratic}
\mu(t,\p) = \log \Bigg(  C \sum _{i=1}^{T}  \; L_{i ,p}* 
	\bigg( \exp \bigg\{- \big|\frac{t-T_{i}}{\Delta t} \big|^2 \bigg\} \bigg) \,  I_{[t',t'']}(t,\p)  
		 +  C \sum _{i=1}^{T} \;\omega_{i,p} \, I_{[t',t'']}(t,\p) \; +1 \Bigg)
\]

\paragraph{M6:}
Here we take into account lifetime phases of convective events in addition to cell's activity:

\begin{equation} \label{eq:part2_mu}
\begin{split}
\mu(t,\p) &= \log \Bigg(  C \sum _{i=1}^{T}  \; L_{i ,p}*\\
	&\bigg( \exp \bigg\{-\big|\frac{t-T_{i}}{\Delta t} \big| \bigg\}  I_{[0,T_{Ch}]}(t) I_{[t',t'']}(t,\p) 
		 + \exp \bigg\{-\big|\frac{t-T_{i}}{\Delta t} \big|^2 \bigg\}  I_{[T_{Ch},T]}(t)  I_{[t',t'']}(t,\p) 
		  \bigg) \nonumber \\
		 &+\; C \sum _{i=1}^{T} \;\omega_{i,p}  I_{[t',t'']}(t,\p) \; +1 \Bigg) 
\end{split}
\end{equation}
 
where $I_{[0,T_{Ch}]}(t)$ and $I_{[T_{Ch},T]}(t)$ are the indicator functions assuming value $1$ if either the predicted time \textit{t} or the observed time $T_{i}$ are in the same lifetime phase (Charging or Mature-Dissipating) and $0$ otherwise whilst $I_{[t',t'']}(t,\p)$ is the indicator function which is $1$ if either the predicted time \textit{t} or the observed time $T_{i}$ are in the activity phase of the time interval $[t',t'']$ and $0$ otherwise. \\

A further model variation, named \emph{memory}, is obtained by changing the way the spatial weights are computed. Instead of considering all $\omega_{i,p}$ over the entire event duration at once, we consider weights along a 30-minutes time window regardless of the cell's activity,  then 
\begin{eqnarray}
\omega^{m}_{i,p}&=&\dfrac{L_{i,p} + \frac{1}{8}\, L_{i,N_{p}} + L_{i-1,p} + \frac{1}{8}\, L_{i-1,N_{p}}}{\sum_{k=1}^{i} \omega_{k,p}}  \label{eq:weight30min_memory}\\
\mbox{for 30-minutes and }\nonumber\\
\omega^{m}_{i,p}&=&\,\dfrac{L_{i,p} + \frac{1}{8}\, L_{i,N_{p}} + L_{i-1,p} + \frac{1}{8}\, L_{i-1,N_{p}} + + L_{i-2,p} + \frac{1}{8}\, L_{i-2,N_{p}}}{\sum_{k=1}^{i} \omega_{k,p}} \label{eq:weight15min_memory} 
\end{eqnarray}
for 15-minutes aggregation, respectively.
This modification let us to  better account for  local features of rainy events. We expect this last variation to be more effective for the May event given its temporal and spatial structure.
Finally, we summarize in Table \ref{tab:model_modifications1} all model modifications we propose. \\

\subsection{The space-time variation}
\label{sec:part2_space_time_effect}

The $w(t,\p)$ component explains the residual spatial variation of rainfall after accounting for that included in the mean representation. In practice, the $\textbf{W}$ component determines a random increase/decrease of the intercept and it describes the overall variance structure. The $w(t,\p)$ is the $(t,\p)$ element of $\boldsymbol{W}$, a separable space-time random field such that $w(t,\p)=T(t)+S(\p)$ where the temporal part is specified as an autoregressive model of order 1: $T(t)=\alpha T(t-1)+\eta(t)$ with $\eta(t)\sim N(0,\sigma^2_\eta)$ and the spatial part is a Gaussian Conditional Autoregressive model (CAR) \citep{Besag:1974}. The CAR model is characterized by a clear link between the conditional and the joint probability distributions and, consequently, we can use the local information to make inference on the random field \textbf{S}  \citep[details can be found for example in ][]{Besag:1974, Cressie:1993, Banerjee:2004}. Formally  the spatial  random field $\mathbf{S}=( S_{1}, \cdots ,S_{p}, \cdots ,S_{n} )$ for all times $t$ has a multivariate normal distribution 

\begin{equation}\label{eq:CAR}
\mathbf{S}\sim MN(\mathbf{0},\sigma^2_{S}(\mathbf{I}-\rho_{S}\mathbf{B})^{-1})
\end{equation}
where $\sigma^2_{S}$ is the process variance, $\B$ is an adjacency matrix such that $b_{ij}=1$ if cell $i$ is a neighbor of cell $j$ and zero otherwise, $\rho_{S}$ is coefficient measuring the strength of spatial interactions and it has to lie in the interval  $(1/m,1)$ with $m$ the maximum number of neighbors in the adopted neighborhood structure. This last condition ensures that \eqref{eq:CAR} is well defined (see for example Chapter 5 of \citet{Banerjee:2004}). \\

\paragraph{\textit{Prior distributions.}}\label{subsec:model_summary}

We suggest the following \textbf{prior distributions} for the models parameters:
 $\alpha \sim N (\mu_{\alpha},\; \sigma_{\alpha}^2)$,
						$\sigma_{\eta}^2 \sim Inv\Gamma(a_{{\eta}},b_{{\eta}})$;
						$\sigma_{S}^2 \sim Inv\Gamma(a_{{S}},b_{{S}})$,
						$\rho_{S} \sim N(0, \sigma^2_\rho)I_{(0,1/m)}$;
						$\sigma^2\sim Inv\Gamma(a_\tau,b_\tau)$ and for 
		\textbf{M2:} $a\sim \Gamma(a_0,b_0)$, $b\sim\Gamma(a_1,b_1)$ and 
		 \textbf{M3:} $b1\sim \Gamma(a_1,b_1)$, $b2\sim\Gamma(a_1,b_1)$.

\begin{table}
\caption{Summary of proposed models.
\label{tab:model_modifications1}} 
\begin{center}
\begin{small}
\begin{tabular}{lrrrr}
\hline
Model modifications&\multicolumn{1}{c}{Velocity}&\multicolumn{1}{c}{Phases}&\multicolumn{1}{c}{Pixel Activity}&\multicolumn{1}{c}{Parameters}\\ 
\hline
M1&-&X&& $\theta =\{\alpha,\tau_{\eta}^2,\tau_{S}^2,\rho_{S},\tau^2 \}$\\
M2&X&X&-&$\theta =\{a,b,\alpha,\tau_{\eta}^2,\tau_{S}^2,\rho_{S},\tau^2 \}$\\
M3&X&X&-&$\theta =\{b_1,b_2,\alpha,\tau_{\eta}^2,\tau_{S}^2,\rho_{S},\tau^2 \}$\\
M4&-&-&X&$\theta =\{\alpha,\tau_{\eta}^2,\tau_{S}^2,\rho_{S},\tau^2 \}$\\ 
M5&-&-&X&$\theta =\{\alpha,\tau_{\eta}^2,\tau_{S}^2,\rho_{S},\tau^2 \}$\\ 
M6&-&X&X&$\theta =\{\alpha,\tau_{\eta}^2,\tau_{S}^2,\rho_{S},\tau^2 \}$\\ 
\hline
M2memory&X&X&-&$\theta =\{a,b,\alpha,\tau_{\eta}^2,\tau_{S}^2,\rho_{S},\tau^2 \}$\\
M3memory&X&X&-&$\theta =\{b_1,b_2,\alpha,\tau_{\eta}^2,\tau_{S}^2,\rho_{S},\tau^2 \}$\\
\hline
\end{tabular}
\end{small}
\end{center}
\end{table}

Notice that the inclusion of a $\delta$ lightning-rain delay could be eventually inserted in the model.\\
In our implementation we set the proposed priors as  $a,b,b_1,b_2 \, \sim \Gamma(0.001,0.001)$, for computational reasons we simulate using precisions instead of variances then $\tau_{\eta}^{2}=\sigma_\eta^{-2}$ and $\tau_{S}^{2}=\sigma_S^{-2}$ $\sim \Gamma(0.001,0.001)$.  $\alpha\sim N(0.5,100)$ and $\rho_{S} \sim U(0, 1/m)$ where \emph{m} is the maximum number of neighbors. Finally, $\tau^{2}=\sigma^{-2}\sim \Gamma(0.001,0.001)$. \\

\paragraph{\textit{Estimation: predictive distribution.}}
The Gaussian framework we adopt simplifies the building of the predictive distribution of the latent process $\Y$.  Consider an unobserved location in space and time $(t_0,\p_0)$, the joint distribution of $(Y(t_0,\p_0),\Y)|\btheta,\mathbf{W}$ is Gaussian as well as the conditional distribution $Y(t_0,\p_0)|\Y,\btheta,\mathbf{W}$. The predictive distribution is obtained integrating over the parameters set and it is analytically intractable. However we can draw samples from it using Monte Carlo methods. We have $N^*$ posterior samples of $\btheta$ and $\mathbf{W}$, for each sample $j$ ($j=1,\ldots,N^*$) we simulate $Y_j(t_0,\p_0)$ from $\pi(Y(t_0,\p_0)|\Y,\btheta_j,\mathbf{W}_j)$ and use it for inferential purposes.

\section{Results}
\label{sec:Results}

\subsection{Estimation of parameters}
\label{sec:part2_parameters_estimation}

\begin{figure}
\subfigure[]{
  \includegraphics[width=0.4\textwidth]{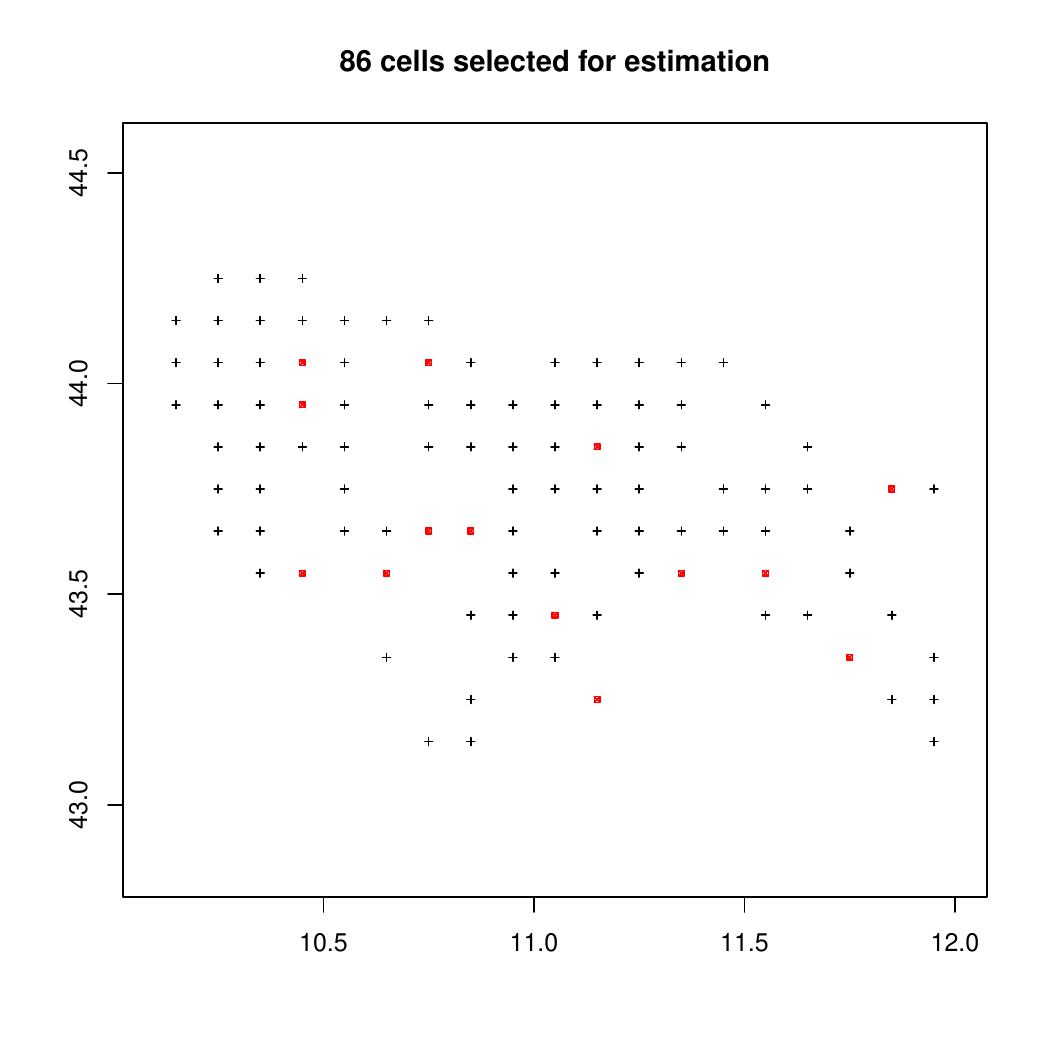}}
\subfigure[]{
  \includegraphics[width=0.4\textwidth]{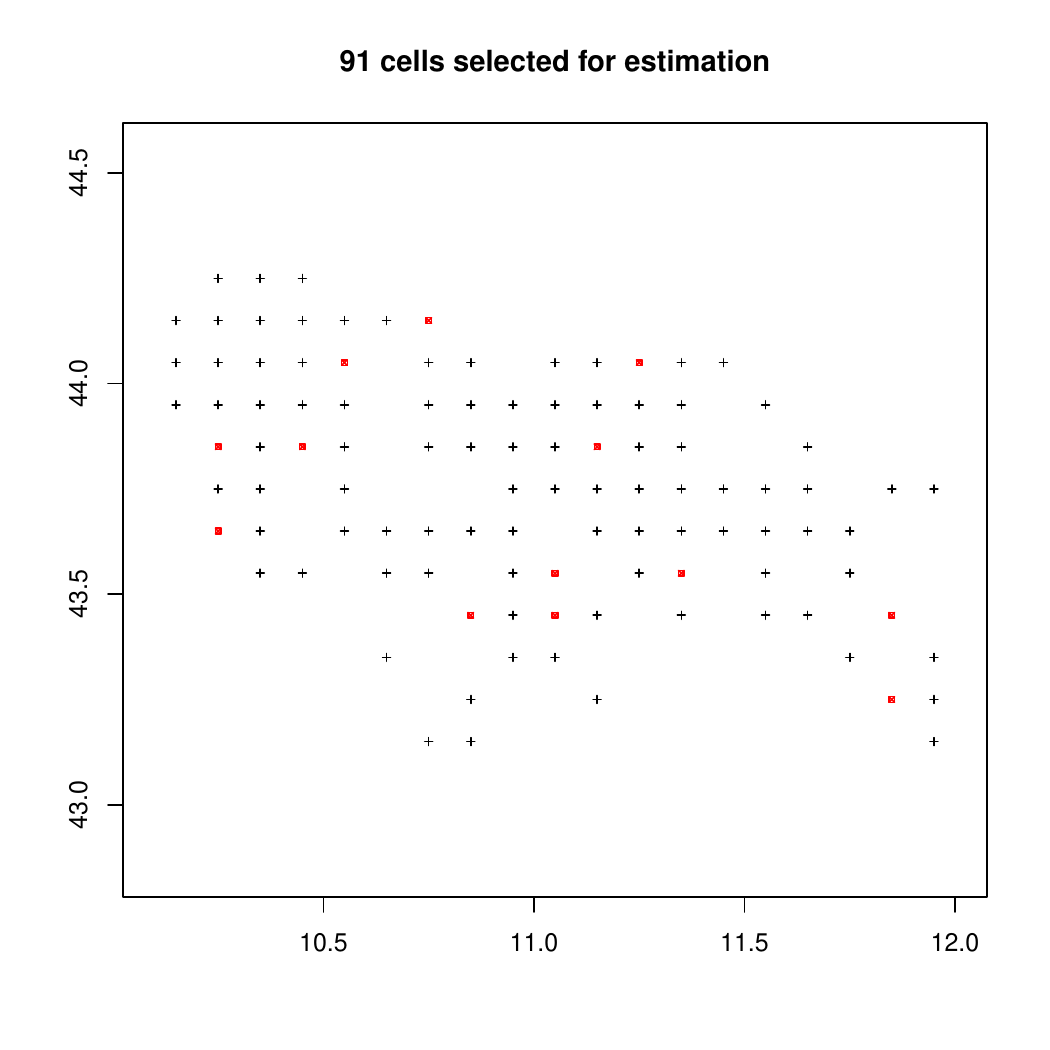}}\\
\subfigure[]{
  \includegraphics[width=0.4\textwidth]{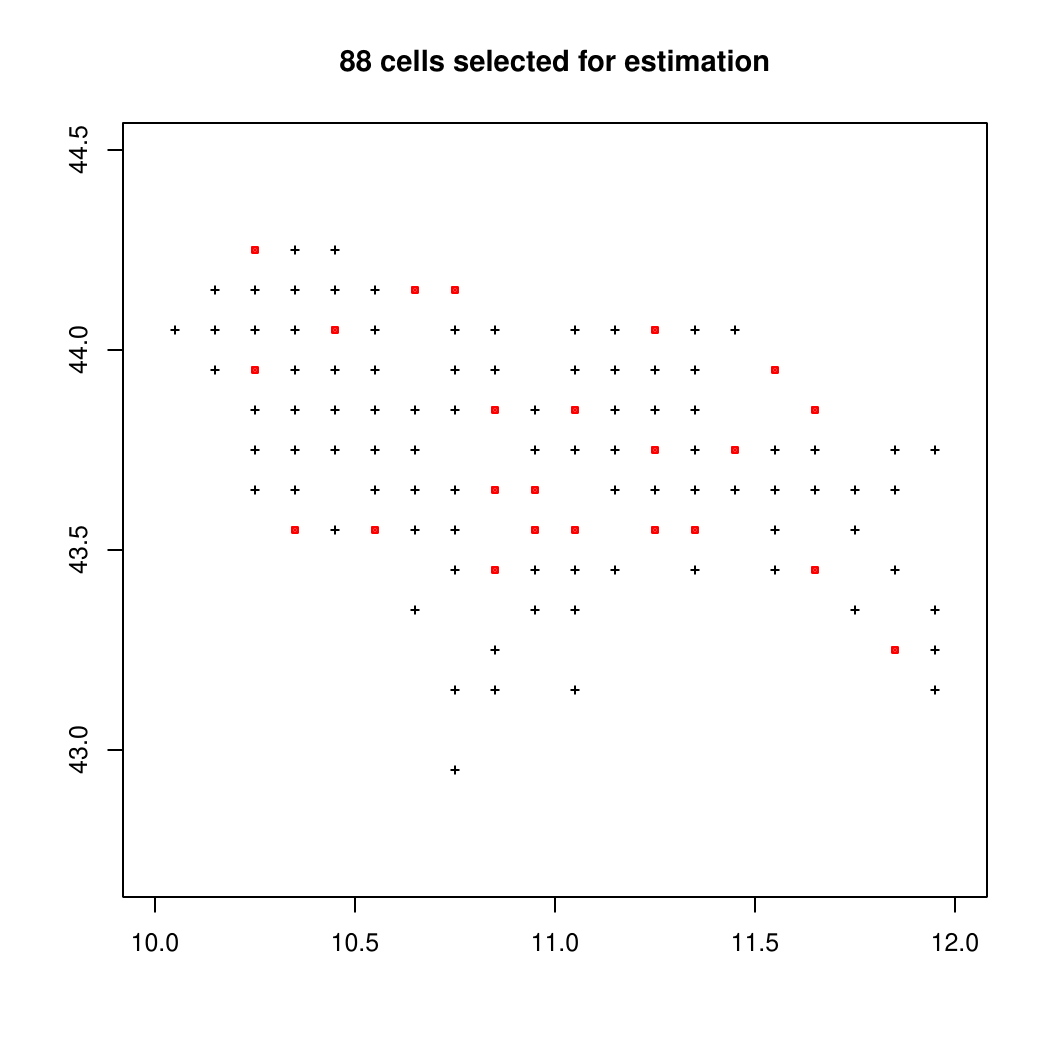}}
\subfigure[]{
  \includegraphics[width=0.4\textwidth]{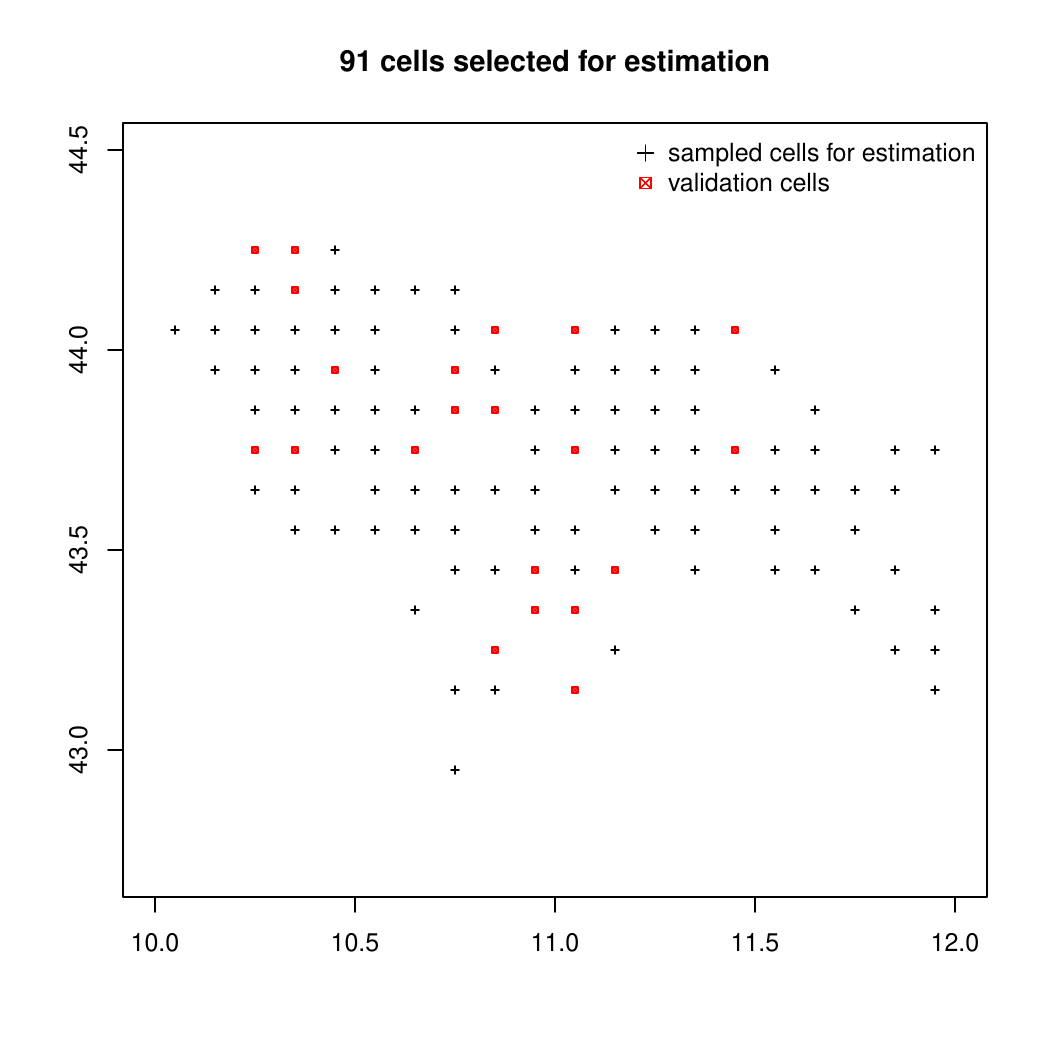}}
\caption{Grid locations  (a)-(b) August 5, 2004 15-minutes and 30-minutes time scales; (c)-(d) May 9, 2006 15-minutes and 30-minutes time scales. In panel (a)-(d) the selected cells for estimation (black crosses) and validation cells (red boxes) are mapped.} 
\label{fig:part2_neighborhood_9May2006}       
\end{figure}

In the definition of the spatial weights and of the spatial dependence we adopt a second order nearest-neighbor structure as shown in Figure \ref{fig:part2_spatial_function}.  A random sample of cells is drawn from the set of grid cells with at least one rain gage $D=\{1, \cdots ,p, \cdots,n\}$ for estimation. The remaining cells of the same sub-grid  are set aside for validation purposes (see Figure \ref{fig:part2_neighborhood_9May2006}). A summary description of the training and validation sets  is reported in Table \ref{tab:part2_riepilogo_dependence}.

\begin{table}
\caption{Number of estimation  (Est.) and validation (Val) cells and maximum number of neighbors (m) in the spatial neighborhood  structure for each case study.}
\label{tab:part2_riepilogo_dependence}
\begin{center}
\begin{tabular}{lrrrr}			
\hline
Case&\multicolumn{1}{c}{n}&\multicolumn{1}{c}{Est(\#cells)}&\multicolumn{1}{c}{Val(\#cells)}&\multicolumn{1}{c}{m}\\
\hline
5Aug2004-15min&$100$&$86$&$14$&$8$\tabularnewline
5Aug2004-30min&$104$&$91$&$13$&$8$\tabularnewline
\hline
9May2006-15min&$111$&$88$&$23$&$7$\tabularnewline
9May2006-30min&$112$&$91$&$21$&$7$\tabularnewline
\hline
\end{tabular}
\end{center}
\end{table}

\begin{figure}
\subfigure[]{
\includegraphics[width=0.6\textwidth]{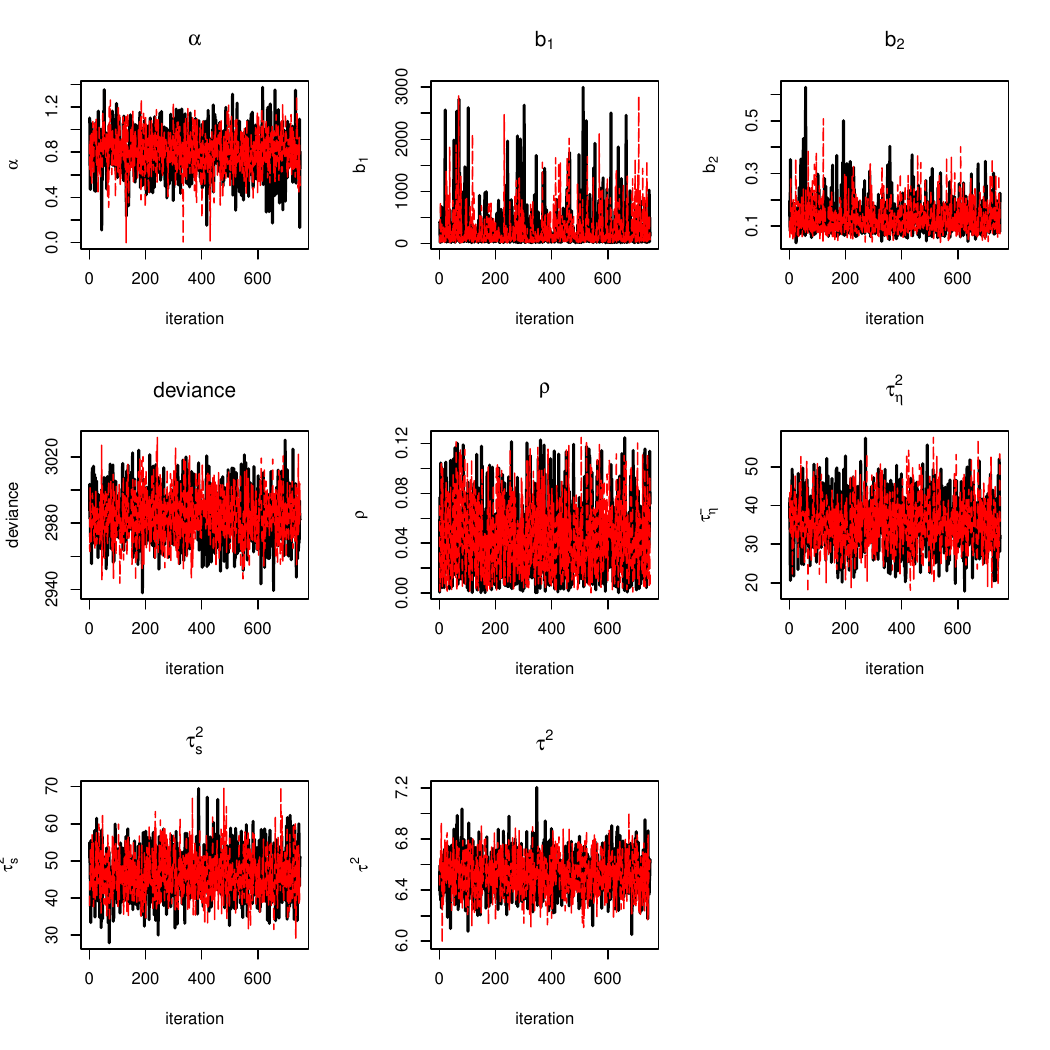}
}
\subfigure[]{
\includegraphics[width=0.6\textwidth]{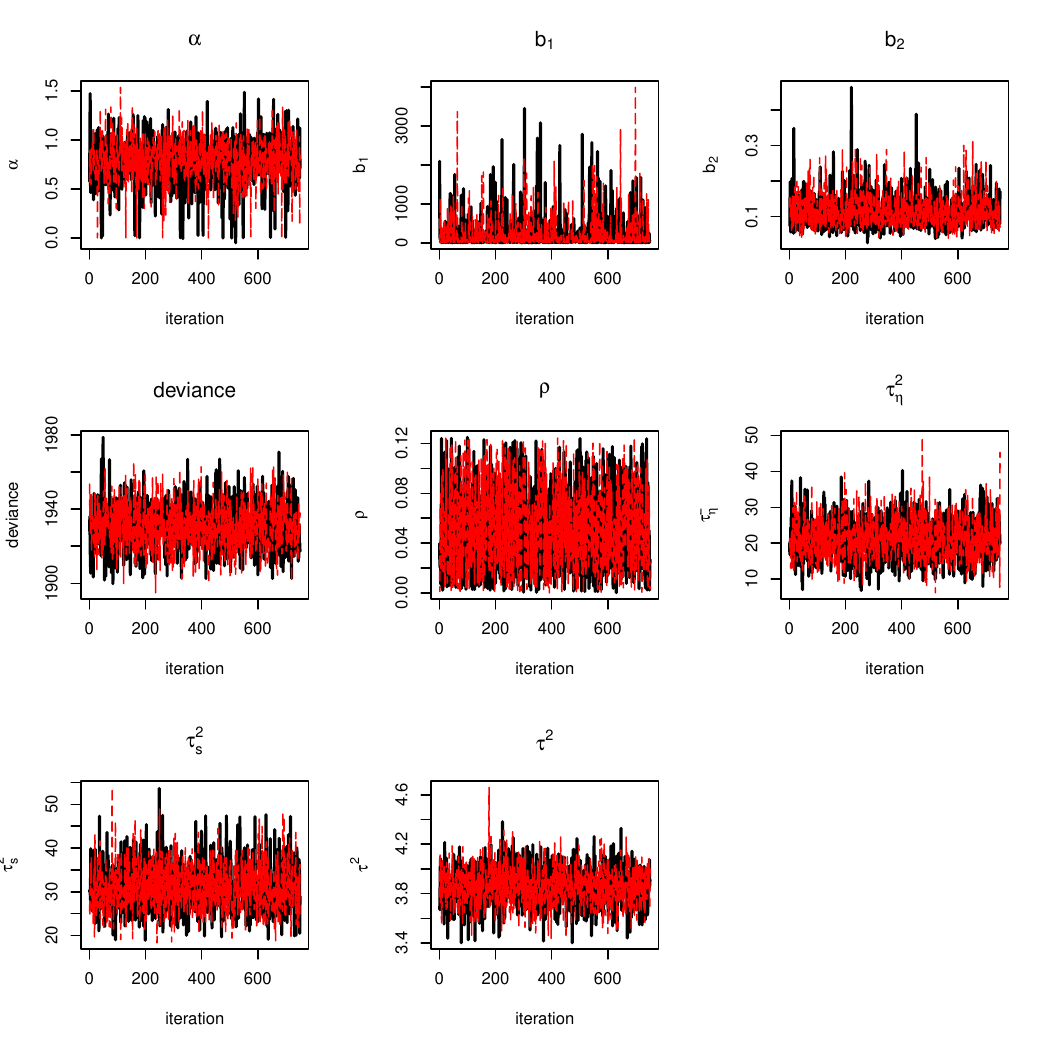}
}
\caption{Trace-plots obtained with 20000 iterations and 2 chains after a burn-in of 5000 and a thin of $ 1/10 $: a)  M3 for August 5, 2004 at 15-minutes time scale, b) M3 for August 5, 2004 at 30-minutes time scale.}
\label{fig:trace_plots_Aug}       
\end{figure}

\begin{figure}
\subfigure[]{
\includegraphics[width=0.6\textwidth]{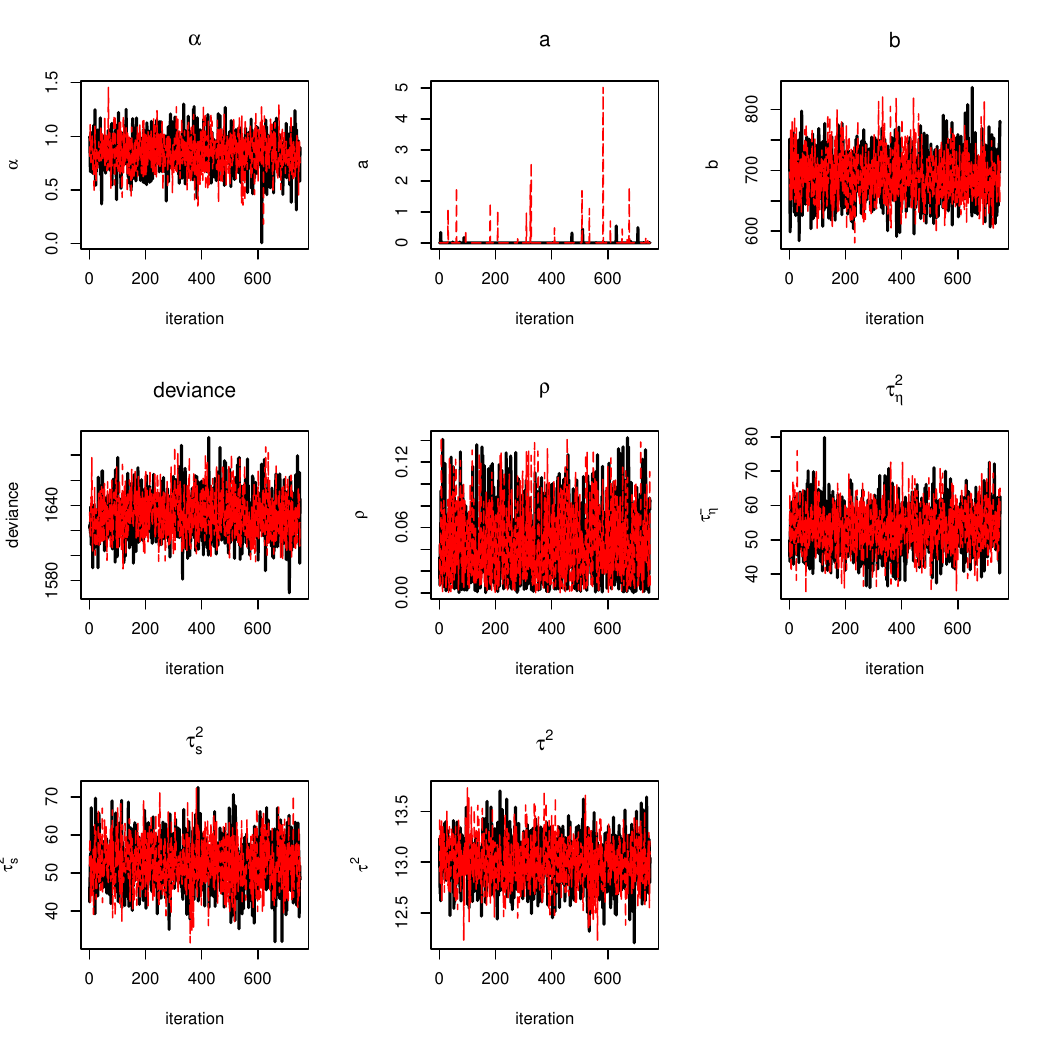}
}
\subfigure[]{
\includegraphics[width=0.6\textwidth]{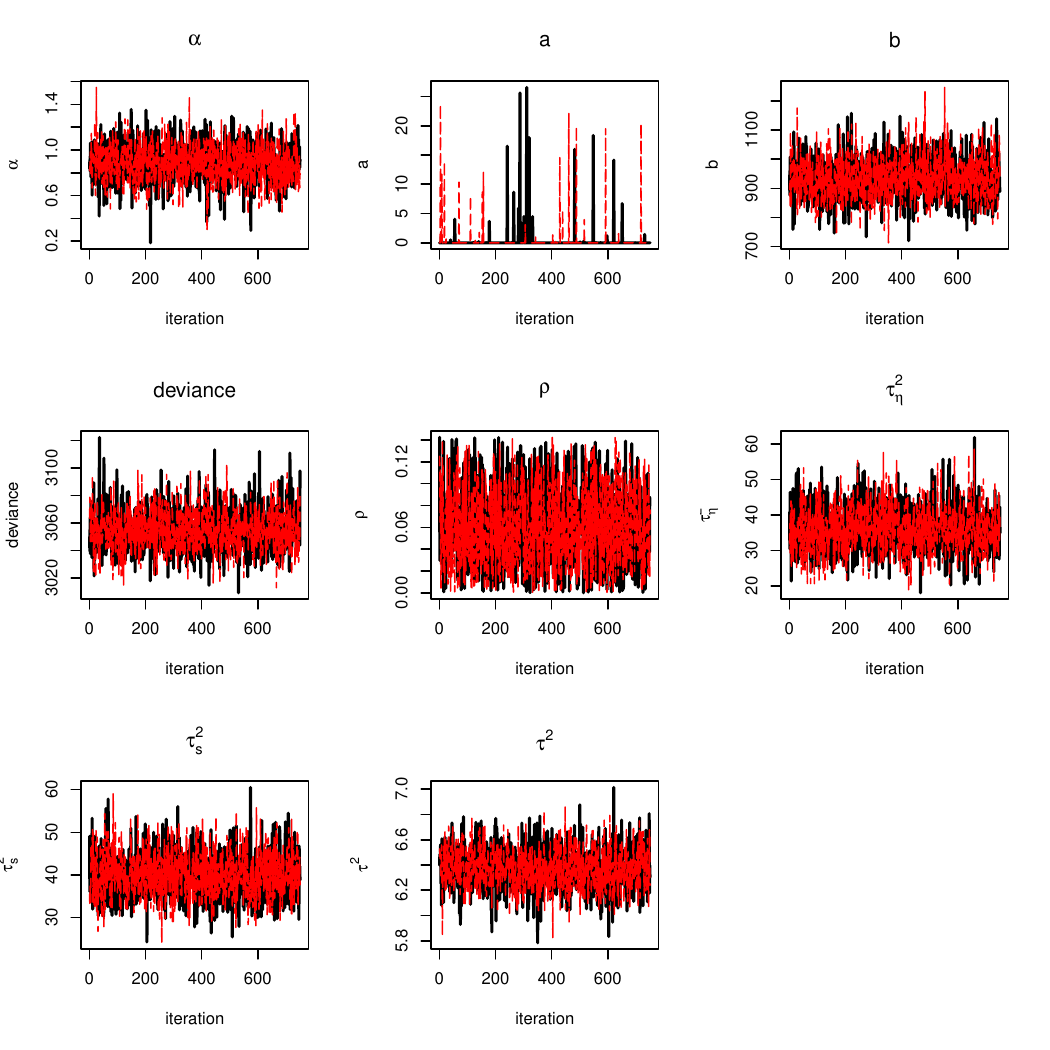}
}
\caption{Trace-plots obtained with 20000 iterations and 2 chains after a burn-in of 5000 and a thin of $ 1/10 $: a) M2memory for May 9, 2006 at 15-minutes time scale, b)  M2 for May 9, 2006 at 30-minutes time scale.}
\label{fig:trace_plots_May}       
\end{figure}

The model estimation is implemented in JAGS \citep{jags} using the package \textit{R2jags} \citep{R2jags:2012} to run the simulation within R. We run two chains with dispersed starting points for 20000 iterations, with a burn-in of 5000 and a thinning of $1/20$ such that we retain the last 1000 iterations of each chain for estimation. Convergence was inspected both graphically and from several summary statistics. Trace-plots of chosen parameters and  summary statistics for the ``best'' models are presented in Figure~\ref{fig:trace_plots_Aug} and ~\ref{fig:trace_plots_May}, and in Table~\ref{tab:PosteriorInference}. Trace-plots and the Potential Scale Reduction Factor $\hat{R}$ proposed by Gelman and Rubin \citep{Gelman:1992} show that the MCMC converges rapidly for all parameters and model setting.

%
\begin{landscape}
\begin{table}
\begin{small}
\caption{MCMC Posterior Inference for best performance cases: M3 for August 5, 2004 at 30-minutes time scale and M2memory for May 9, 2006 at 15-minutes time scale.\label{tab:PosteriorInference}} 
\begin{center}
\begin{tabular}{lrrrrrrrrr}
\hline
\multicolumn{1}{l}{Aug 5, 2004 M3 30-min}&\multicolumn{1}{c}{mean}&\multicolumn{1}{c}{sd}&\multicolumn{1}{c}{2.5\%}&\multicolumn{1}{c}{25\%}&\multicolumn{1}{c}{50\%}&\multicolumn{1}{c}{75\%}&\multicolumn{1}{c}{97.5\%}&\multicolumn{1}{c}{$\widehat{R}$}&\multicolumn{1}{c}{n.eff}\tabularnewline
\hline
$\alpha$&$   0.789$&$  0.226$&$   0.268$&$   0.661$&$   0.802$&$   0.934$&$   1.200$&$1.004$&$ 670$\tabularnewline
$b_1$&$ 233.694$&$424.799$&$  10.733$&$  20.652$&$  62.987$&$ 224.452$&$1504.153$&$1.013$&$ 160$\tabularnewline
$b_2$&$   0.115$&$  0.046$&$   0.053$&$   0.082$&$   0.106$&$   0.138$&$   0.236$&$1.001$&$1500$\tabularnewline
deviance&$1930.806$&$ 12.225$&$1907.488$&$1922.569$&$1930.356$&$1938.901$&$1956.441$&$1.005$&$ 350$\tabularnewline
$\rho$&$   0.051$&$  0.033$&$   0.003$&$   0.023$&$   0.048$&$   0.077$&$   0.119$&$1.000$&$1500$\tabularnewline
$\tau^2_\eta$&$  21.281$&$  5.812$&$  10.824$&$  17.230$&$  21.092$&$  25.060$&$  33.309$&$1.004$&$ 540$\tabularnewline
$\tau^2_s$&$  31.049$&$  5.427$&$  21.864$&$  27.185$&$  30.548$&$  34.568$&$  42.536$&$1.002$&$1300$\tabularnewline
$\tau^2_y$&$   3.855$&$  0.159$&$   3.556$&$   3.744$&$   3.857$&$   3.960$&$   4.165$&$1.003$&$ 540$\tabularnewline
\hline
\multicolumn{1}{l}{May 9, 2006 M2memory 15-min}&\multicolumn{1}{c}{mean}&\multicolumn{1}{c}{sd}&\multicolumn{1}{c}{2.5\%}&\multicolumn{1}{c}{25\%}&\multicolumn{1}{c}{50\%}&\multicolumn{1}{c}{75\%}&\multicolumn{1}{c}{97.5\%}&\multicolumn{1}{c}{$\widehat{R}$}&\multicolumn{1}{c}{n.eff}\tabularnewline
\hline
a&$   0.017$&$ 0.187$&$   0.000$&$   0.000$&$   0.000$&$   0.000$&$   0.000$&$1.035$&$ 550$\tabularnewline
$\alpha$&$   0.851$&$ 0.165$&$   0.515$&$   0.743$&$   0.853$&$   0.962$&$   1.164$&$1.046$&$1500$\tabularnewline
b&$ 690.969$&$39.417$&$ 619.094$&$ 662.688$&$ 689.590$&$ 715.333$&$ 771.675$&$1.000$&$1500$\tabularnewline
deviance&$1632.487$&$17.594$&$1600.403$&$1619.470$&$1632.540$&$1643.701$&$1668.750$&$1.000$&$1500$\tabularnewline
$\rho$&$   0.046$&$ 0.033$&$   0.002$&$   0.019$&$   0.040$&$   0.068$&$   0.122$&$1.005$&$1500$\tabularnewline
$\tau^2_\eta$&$  53.367$&$ 6.718$&$  40.625$&$  48.499$&$  53.318$&$  57.861$&$  66.602$&$1.001$&$1500$\tabularnewline
$\tau^2_s$&$  52.385$&$ 6.343$&$  40.560$&$  47.972$&$  52.230$&$  56.664$&$  65.288$&$1.000$&$1500$\tabularnewline
$\tau^2_y$&$  13.001$&$ 0.235$&$  12.548$&$  12.836$&$  13.002$&$  13.160$&$  13.449$&$1.000$&$1500$\tabularnewline
\hline
\end{tabular}
\end{center}
\end{small}
\end{table}
\end{landscape}

\subsection{Evaluation of predictions}
\label{subsec:evaluation}
We use two categories of indexes to evaluate the predictive performance of our modeling approach: 1) for the first category, we compute the Deviance Information Criterion (DIC) for evaluating model fit and the Empirical Coverage (EC) of $90\%$ credible intervals for evaluating the credibility of simulation results; 2) for the second category, we adopt the same validation scheme used  for meteorological variables (see for example \citet{Murphy:1985} or \citet{Thornes:2001}) to analyze the quality of predictions over the validation set. In particular, we use Root Mean Square Errors (RMSE) to measure the mean error when considering predictions and observations as continuous variables;  Probability of Hits on Total (POHT), Probability of Detection (POD) and Probability of False Detection (POFD) when classifying events as dichotomous variables into rain/no rain, that is $[0,0.2),\;[0.2,\infty)$ and, finally, Heidke skill score (HSS) always for the same two categories.   
To facilitate the understanding of indexes role we report their definition in Table \ref {tab:part2_POD_theory} following \citet{Far:2009} notation. POHT is a measure of accuracy since it is the percentage of cases correctly predicted in both classes \textit{Rain} and \textit{No Rain}; POD is the rate  correctly predicted  \textit{Rain} events  and POFD, also known as False Alarm Ratio (FAR), is the rate of events mistakenly  predicted as rainy. 
It is worth considering POHT when predictions of \textit{Rain} and \textit{No Rain} events have, as in the present study, the same relevance. On the other hand, POD and POFD  are focused on the evaluation of predictions in the \textit{Rain} class. Nevertheless, a joint analysis of these indexes is generally recommended \citep{Far:2009} to fully understand model's performances. Finally, HSS measures the accuracy of the model in predicting the correct category with respect to results due to random chance. HSS ranges from $-\infty$ to $1$: $0$ indicates no skill, $1$ is the perfect score. 

\begin{table}
\begin{small}
\caption{Calculation of Probability of Hits on Total (POHT), Probability of Detection (POD), Probability of False Detection (POFD or FAR) and Heidke skill score (HSS).}
\label{tab:part2_POD_theory}
  \begin{center}
    \begin{tabular}{llcc|l}
    \hline
    \multicolumn{1}{l}{}&\multicolumn{1}{l}{}&\multicolumn{2}{c}{\textbf{Observed}}\tabularnewline
    \cline{3-4}
   \multicolumn{1}{l}{}&\multicolumn{1}{l}{}&\multicolumn{1}{c}{\textit{Rain}}&\multicolumn{1}{c}{\textit{No Rain}}&\multicolumn{1}{c}{}\\
    \hline
    \textbf{Predicted} & \textit{Rain}   & \texttt{a} & \texttt{b} & \texttt{a+b}\\
                                            &\textit{No Rain} & \texttt{c} & \texttt{d} & \texttt{c+d}\\
    \hline
              &      & \texttt{a+c} & \texttt{b+d} & \texttt{n}\\
    \hline\\
  \multicolumn{1}{l}{Indexes}&\multicolumn{1}{c}{Acronym}&\multicolumn{1}{c}{Formula}\\
  \hline
  Probability of Hits on Total & POHT & $(a+d)/n$ \\
  Probability of Detection & POD & $a/(a+c)$\\
  Probability of False Detection & POFD & $b/(b+d)$\\
  Heidke Skill Score &HSS& $\frac{(a+d)-[(a+c)(a+b)+(d+c)(d+b)]/n}{n -[(a+c)(a+b)+(d+c)(d+b)]/n}$\\
  \hline
    \end{tabular}
  \end{center}
 \end{small}
 \end{table}


The values of DIC, POHT, POD, POFD, RMSE and HSS obtained from predictions of August 5, 2004 and May 9, 2006 are reported in Table \ref{tab:PredictionsEvaluation_5Aug2004} and \ref{tab:PredictionsEvaluation_9May2006}, respectively. 


\begin{table}
\begin{footnotesize}
\caption{Evaluation of MCMC estimates for the events August 5, 2004 15-minutes and August 5, 2004 30-minutes (``best''  model in bold).  \label{tab:PredictionsEvaluation_5Aug2004}} 
\begin{center}
\begin{tabular}{lrrrrrrrr}
\hline
5-Aug-2004-15min&\multicolumn{1}{c}{DIC}&\multicolumn{1}{c}{EC(\%)}&\multicolumn{1}{c}{POHT(\%)}&\multicolumn{1}{c}{POD(\%)}&\multicolumn{1}{c}{POFD(\%)}&\multicolumn{1}{c}{RMSE(mm)}&\multicolumn{1}{c}{HSS}\tabularnewline
\hline
M1&$3606.7$&$92.9$&$73.4$&$84.9$&$33.2$&$0.596$&$0.474$\tabularnewline
M2&$3115.9$&$93.2$&$75.0$&$82.2$&$29.2$&$0.546$&$0.496$\tabularnewline
\textbf{M3}&$\textbf{3097.4}$&$\textbf{93.3}$&$\mathbf{75.4}$&$\mathbf{83.2}$&$\mathbf{29.2}$&$\textbf{0.543}$&$\textbf{0.505\textbf{}}$\tabularnewline
M4&$3566.3$&$92.8$&$74.4$&$84.9$&$31.7$&$0.604$&$0.491$\tabularnewline
M5&$3487.9$&$92.9$&$74.8$&$80.5$&${28.5}$&$0.595$&$0.490$\tabularnewline
M6&$3613.1$&$92.9$&$73.8$&${84.9}$&$32.6$&$0.606$&$0.481$\tabularnewline
\hline

5-Aug-2004-30min& & & & & & & &\tabularnewline
\hline
M1&$2165.3$&$92.6$&$75.6$&${87.8}$&$33.1$&$0.575$&$0.522$\tabularnewline
M2&$2035.1$&$92.9$&$76.9$&$85.7$&$29.4$&${0.524}$&$0.543$\tabularnewline
\textbf{M3}&$\textbf{2005.3}$&$\textbf{92.6}$&$\textbf{79.1}$&$\textbf{87.8}$&$\textbf{27.2}$&$\textbf{0.529}$&$\textbf{0.585}$\tabularnewline
M4&$2227.3$&${92.4}$&$77.8$&$78.6$&$22.8$&$0.567$&$0.550$\tabularnewline
M5&$2253.4$&$92.5$&${80.8}$&$82.7$&${20.6}$&$0.548$&${0.611}$\tabularnewline
M6&$2307.3$&$92.7$&$77.4$&$79.6$&$24.3$&$0.578$&$0.543$\tabularnewline
\hline
\end{tabular}
\end{center}
\end{footnotesize}
\end{table}
%

%
\begin{table}
\begin{footnotesize}
\caption{Evaluation of MCMC estimates for the events May 9, 2006 15-minutes and May 9, 2006 30-minutes (``best''  model in bold).  \label{tab:PredictionsEvaluation_9May2006}} 
\begin{center}
\begin{tabular}{lrrrrrrrr}
\hline
9-May-2006-15min&\multicolumn{1}{c}{DIC}&\multicolumn{1}{c}{EC(\%)}&\multicolumn{1}{c}{POHT(\%)}&\multicolumn{1}{c}{POD(\%)}&\multicolumn{1}{c}{POFD(\%)}&\multicolumn{1}{c}{RMSE(mm)}&\multicolumn{1}{c}{HSS}\tabularnewline
\hline
M1&$2160.7$&$91.4$&$74.2$&$56.3$&$20.8$&$0.424$&$0.319$\\
\textbf{M2}&$\textbf{2158.4}$&$\textbf{91.3}$&$\textbf{75.0}$&$\textbf{56.6}$&$\textbf{19.9}$&$\textbf{0.424}$&$\textbf{0.333}$\tabularnewline
M3&$2168.0$&$91.4$&$74.9$&$ 57.5$&$20.2$&${0.419}$&${0.336}$\tabularnewline
M4&$3803.4$&$92.7$&$69.6$&${58.4}$&$27.3$&$0.448$&$0.257$\tabularnewline
M5&$3759.5$&$92.6$&$69.6$&${58.4}$&$27.3$&$0.448$&$0.258$\\
M6&$2543.0$&$91.6$&$73.0$&$56.0$&$22.3$&$0.435$&$0.298$\tabularnewline
\hline

9-May-2006-30min& & & & & & &\\
\hline
M1&$3203.9$&${91.0}$&${69.9}$&${67.9}$&${29.2}$&$0.560$&${0.353}$\tabularnewline
\textbf{M2}&$\textbf{3167.8}$&$\textbf{91.4}$&$\textbf{69.2}$&$\textbf{66.5}$&$\textbf{29.6}$&$\textbf{0.557}$&$\textbf{0.337}$\tabularnewline
M3&${3173.6}$&${91.3}$&${68.5}$&${65.1}$&$30.0$&${0.564}$&${0.321}$\tabularnewline
M4&$4041.2$&$91.3$&$66.4$&$64.7$&$32.9$&$0.641$&$0.286$\tabularnewline
M5&$4017.4$&$91.5$&$65.1$&$67.0$&$35.7$&$0.632$&$0.276$\tabularnewline
M6&$3994.5$&$91.3$&$66.0$&$67.4$&$34.7$&$0.626$&$0.290$\tabularnewline
\hline
\end{tabular}
\end{center}
\end{footnotesize}
\end{table}

Recall  from Section \ref{subsec:StudyEvents} that the two case study differ substantially in terms of both spatial extent and time duration. Furthermore, they have a different physical structure since the August event  is formed by two storms merging near the Appenini Mountain: one is a propagating system moving from west to east and the other is a thermo-convective event generated by the orography, whilst the  May event is unique and propagates linearly from west to east .\\
The empirical coverage of all models is close to the nominal value of  $90\%$, suggesting a very good accuracy level of the proposed  modeling approach. 
Through  the DIC we  select the best fitting model for each situation, which is M3  for both time scale in the  August 2004 event and M2 for the May 2006 event. In practice, adding the velocity of propagation into the model (M2, M3) improves the predictive capacity.  The simpler M1 model for the May event at 30-minutes time scale  returns a slightly larger DIC but an equivalent and some times better performance in terms of  verification indexes, suggesting that in this situation the inclusion of the storm propagation speed in the model does not significantly improve its predictive capacity.  The use of two different values for $V$, one  for the Charging phase and the other for the Mature-Dissipating phase is useful only for the August 2004 event, suggesting that this differentiation is relevant for the prediction and analysis of large storms. Furthermore, model modifications M4, M5 and M6 (based  on the relation between cells with lightning activity) do not return better predictions.  

The RMSE ranges from $0.419$ mm up to $0.614$ mm that in absolute terms suggests an average error of  the same magnitude as the rain gages error. In relative terms we can evaluate this error in relation to the maximum recorded  in each situation as a measure of event size. Considering only the best models performances again we appreciate  a better result with larger events with errors ranging from $0.9\%$ (August at 30-minutes) to $1.6 \%$ (August at 15-minutes), while in the May event errors range from $2.8\%$ (15-minutes) to $3.2\%$ (30-minutes). HSS suggests a reasonable performance of the model for the largest event of August and a poor overall performance for the smaller, more complex, May storm. A deeper analysis of the performance indexes (POD, POFD,  POHT) highlights the model difficulty in predicting extreme  rain amounts (zero and peaks). One possible reason for this behavior is the chosen spatial structure that takes into account eight cells around each predicting cell for a total  area of $30\times30$ km$^2$. The latter  maybe a too large area when a high rain rate is recorded in one cell. Convective systems can show very different rain rates during their development. In particular, the larger the rain rate in a cell, the smaller the cells cluster with non zero rainfall. 
In view of the last consideration we estimated the same models with a smaller neighborhood structure (maximum number of neighbors 4) but no consistent improvement was obtained. Then we applied the spatial weights modifications given in \eqref{eq:weight30min_memory} and \eqref{eq:weight15min_memory} to the first three best performing models for both events, the best models results  are shown in Table \ref{tab:VarianteMemory}. The modification of spatial weights improves the model predictive performance for the  May 9, 2006 event whereas it seems to return no clear  advantages for the  Aug 5, 2004 event. 

\begin{table}
\begin{footnotesize}
\caption{Comparison of best performing models and their version with spatial weights given in \eqref{eq:weight30min_memory} and \eqref{eq:weight15min_memory}.  \label{tab:VarianteMemory}} 
\begin{center}
\begin{tabular}{lrrrrrrrr}
\hline
\multicolumn{1}{c}{}&\multicolumn{1}{c}{DIC}&\multicolumn{1}{c}{EC(\%)}&\multicolumn{1}{c}{POHT(\%)}&\multicolumn{1}{c}{POD(\%)}&\multicolumn{1}{c}{POFD(\%)}&\multicolumn{1}{c}{RMSE(mm)}&\multicolumn{1}{c}{HSS}\tabularnewline
\hline
5-Aug-2004& & & & & & &\\
\hline
{M3memory-15min}&${3105.2}$&${93.5}$&${74.6}$&${84.9}$&${31.3}$&${0.536}$&${0.495}$\tabularnewline
\hline
{M3memory-30min}&${1987.3}$&${92.6}$&${77.8}$&${87.8}$&${31.7}$&${0.531}$&${0.561}$\tabularnewline
\hline
\hline
9-May-2006& & & & & & &\\
\hline
\textbf{M2memory-15min}&$\textbf{1787.4}$&$\textbf{91.1}$&$\textbf{76.7}$&$\textbf{56}$&$\textbf{17.6}$&$\textbf{0.420}$&$\textbf{0.360}$\\
\hline
{M3memory-30min}&${2940.0}$&${91.4}$&${69.9}$&${66.1}$&${28.4}$&${0.545}$&${0.346}$\\
\hline
\end{tabular}
\end{center}
\end{footnotesize}
\end{table}

Finally, we select the best performance models including those with \textit{memory} variant: evaluation of MCMC estimates are reported in Table \ref{tab:PredictionsEvaluation_best}.

\begin{table}
	\begin{footnotesize}
	\caption{Evaluation of MCMC estimates for the events 5-Aug-2004-15/30min and 9-May-15/30min, Only the best models are reported.  \label{tab:PredictionsEvaluation_best}} 
	\begin{center}
		\begin{tabular}{lrrrrrrrr}
			\hline
			\multicolumn{1}{c}{}&\multicolumn{1}{c}{DIC}&\multicolumn{1}{c}{EC(\%)}&\multicolumn{1}{c}{POHT(\%)}&\multicolumn{1}{c}{POD(\%)}&\multicolumn{1}{c}{POFD(\%)}&\multicolumn{1}{c}{RMSE(mm)}&\multicolumn{1}{c}{HSS}\tabularnewline
			\hline
			\hline
			5-Aug-2004& & & & & & &\\
			\hline
			15-min: \textbf{M3}&$3097.4$&$93.3$&$75.4$&$83.2$&$29.2$&$0.543$&$0.505$\tabularnewline
			30-min: \textbf{M3}&$2005.3$&$92.6$&$79.1$&$87.8$&$27.2$&$0.529$&$0.585$\tabularnewline
			\hline
			9-May-2006& & & & & & &\\
			\hline
			15-min: \textbf{M2memory}&$1787.4$&$91.1$&$76.7$&$56$&$17.6$&$0.420$&$0.360$\\
			30-min: \textbf{M2}&$3167.8$&$91.4$&$69.2$&$66.5$&$29.6$&$0.557$&$0.337$\tabularnewline
			\hline			
		\end{tabular}
	\end{center}
	\end{footnotesize}
\end{table}

\begin{figure}
	\begin{center}
		\subfigure[August best, corr$=0.92$]{
			\includegraphics[width=0.4\textwidth]{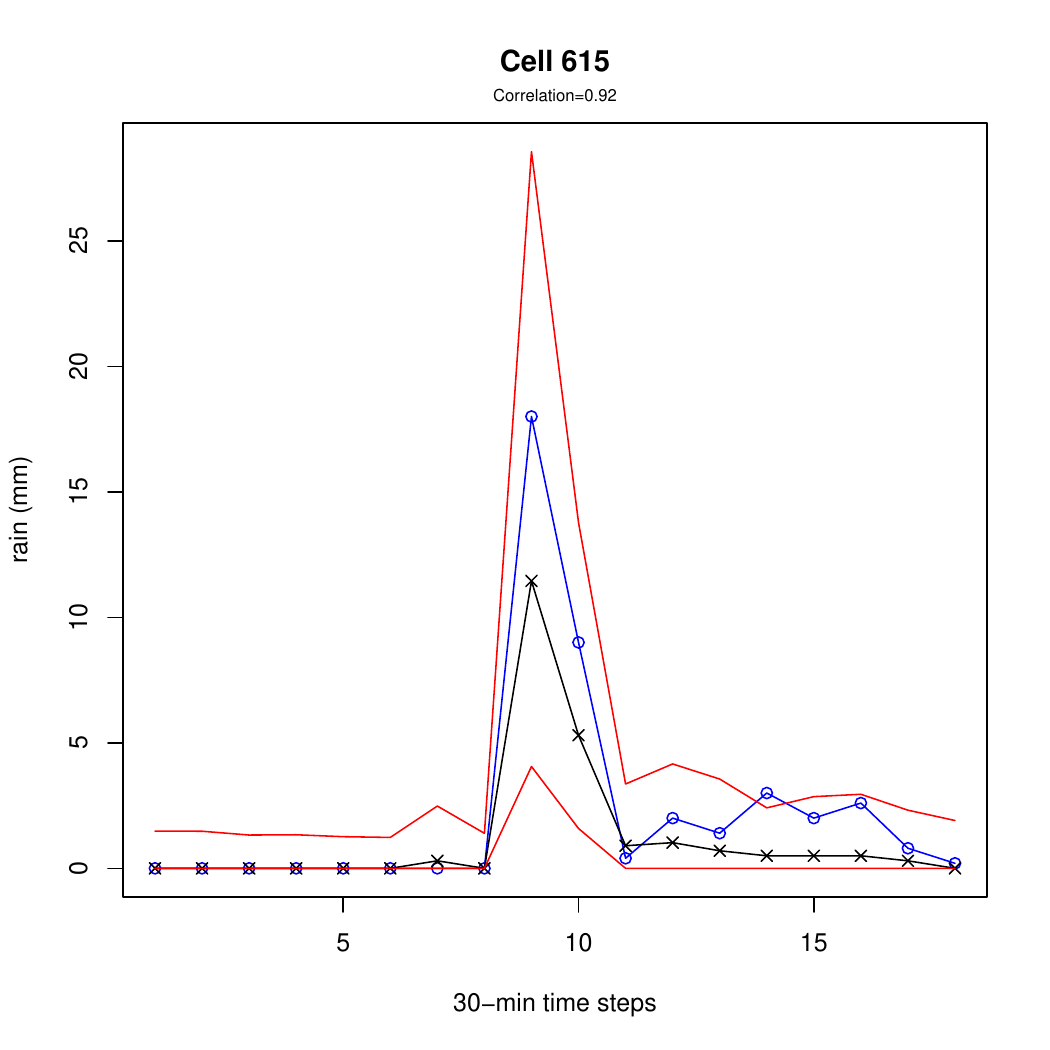}
		}
		\subfigure[August worst, corr$=0.28$]{
			\includegraphics[width=0.4\textwidth]{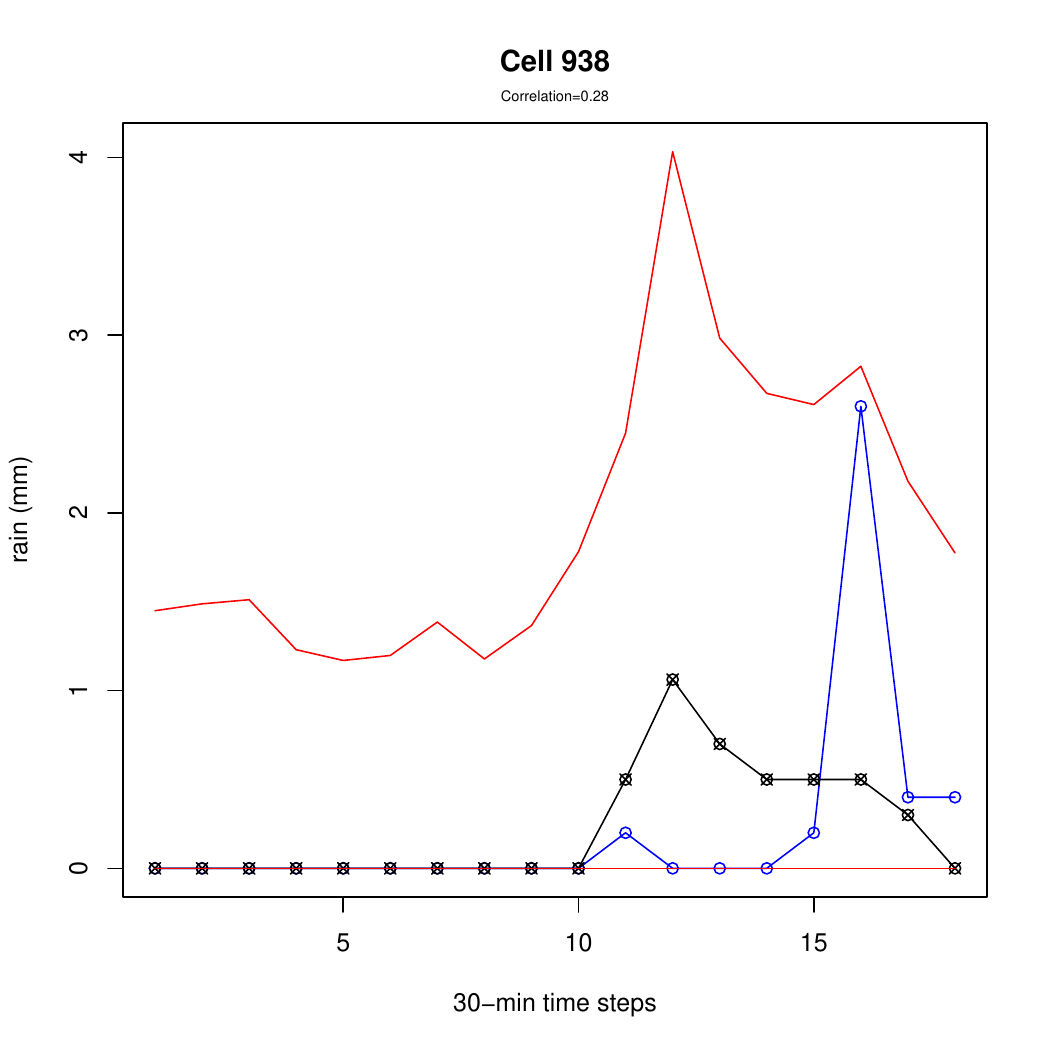}
		}\\
		\subfigure[May best, corr$=0.80$]{
			\includegraphics[width=0.4\textwidth]{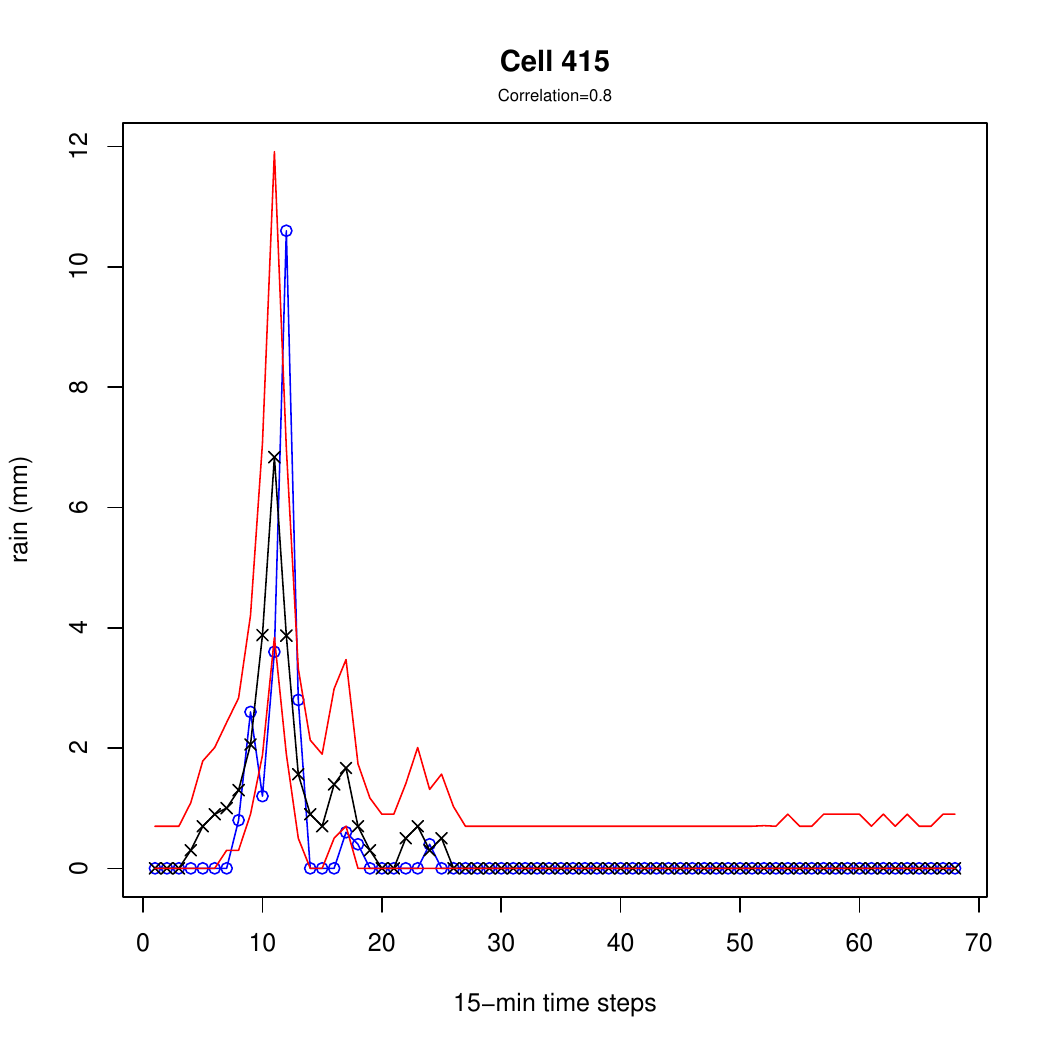}
		}
		\subfigure[May worst, corr$=0.23$]{
			\includegraphics[width=0.4\textwidth]{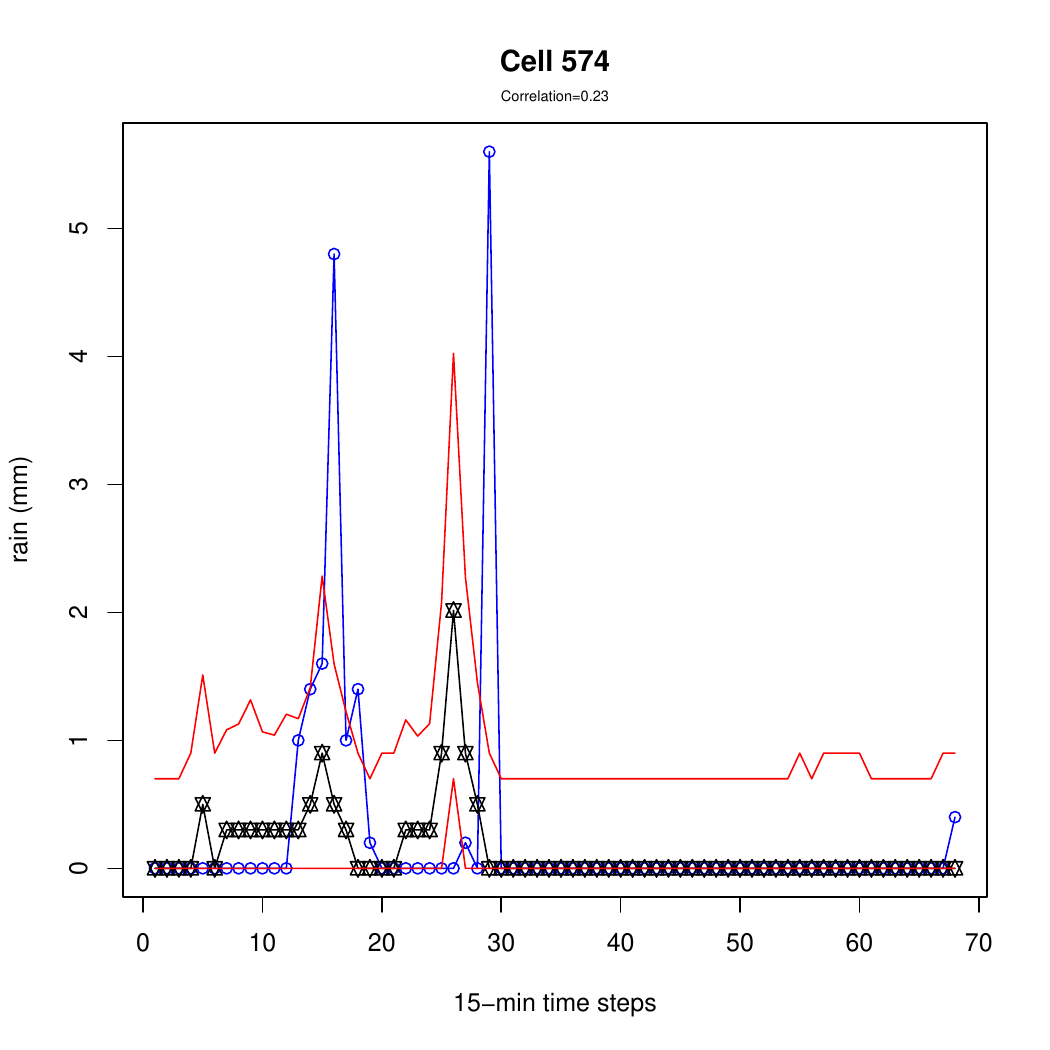}
		}
		\caption{Time series of rainfall predicted values at selected validation cells  for M3 Aug 5, 2004 at  30-minutes time scale ((a) best and (b) worst) and M2memory May 9, 2006 at 15-minutes time scale ((c) best and (d) worst). Predicted values in black line, observed values blue line, 90\% credibility intervals red line.} 
		\label{fig:predictions_timeseries}
	\end{center}
\end{figure}

Model M3 shows an acceptable performance for both time scales in the August event. Only a tendency to underestimate the number of zero observation is highlighted by the POFD values. Time dynamic is well captured by the best models almost everywhere with few exceptions corresponding to extreme recordings. As an example, we report  ``best'' and ``worst'' estimated values, the predicted (black-crossed) series together with the observed (blue line) ones by M3 in August at 30-minutes (panel (a) of Figure \ref{fig:predictions_timeseries}) and by M2memory in May at 15-minutes (panel (b) of Figure \ref{fig:predictions_timeseries}) together with the $90\%$ credibility intervals (red lines). In the plots the correlation between observed and predicted values is reported.  The worst performances are always shown during the smaller event (May) regardless the time scale. The main issue seems to be  the prediction of sudden changes in the series, while the prediction of other peaks is almost always correct in both their location (in space and time) and size.

In summary  results obtained varying the neighborhood dependence structure or allowing for one time step delay (not shown) in the lightning-rainfall relation show that these structure changes seems to be not clearly influential on estimates quality.
On the other hand, some improvement is observed when limiting the effect of neighborhood structure to the 30-minutes before the observed time, in particular a DIC reduction and a slight  general improvement for May 9, 2006 event (see Table \ref{tab:VarianteMemory}). This improvement is due to the fact that this modification introduces a limited  \emph{memory} on the spatial relation helping to account for the  slow movement of the entire system along its 16 hours of duration. On the contrary, an unlimited version of spatial weights, such as in M3 is best performing for more rapid August event (9 hours of duration). \\

\begin{figure}
  \begin{center}
  \subfigure[]{
    	\includegraphics[width=0.5\textwidth]{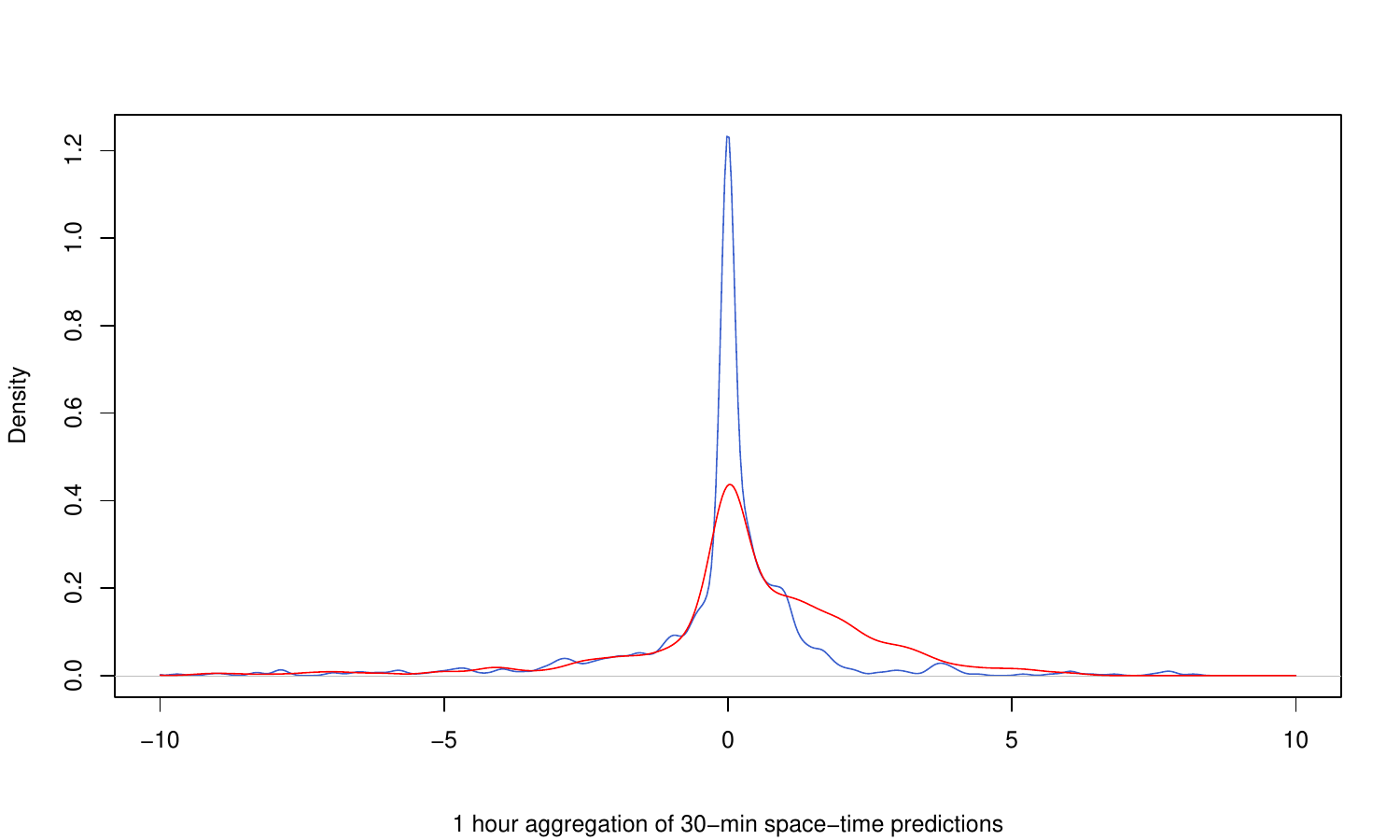}
  }\\ 
  \subfigure[]{
    \includegraphics[width=0.4\textwidth]{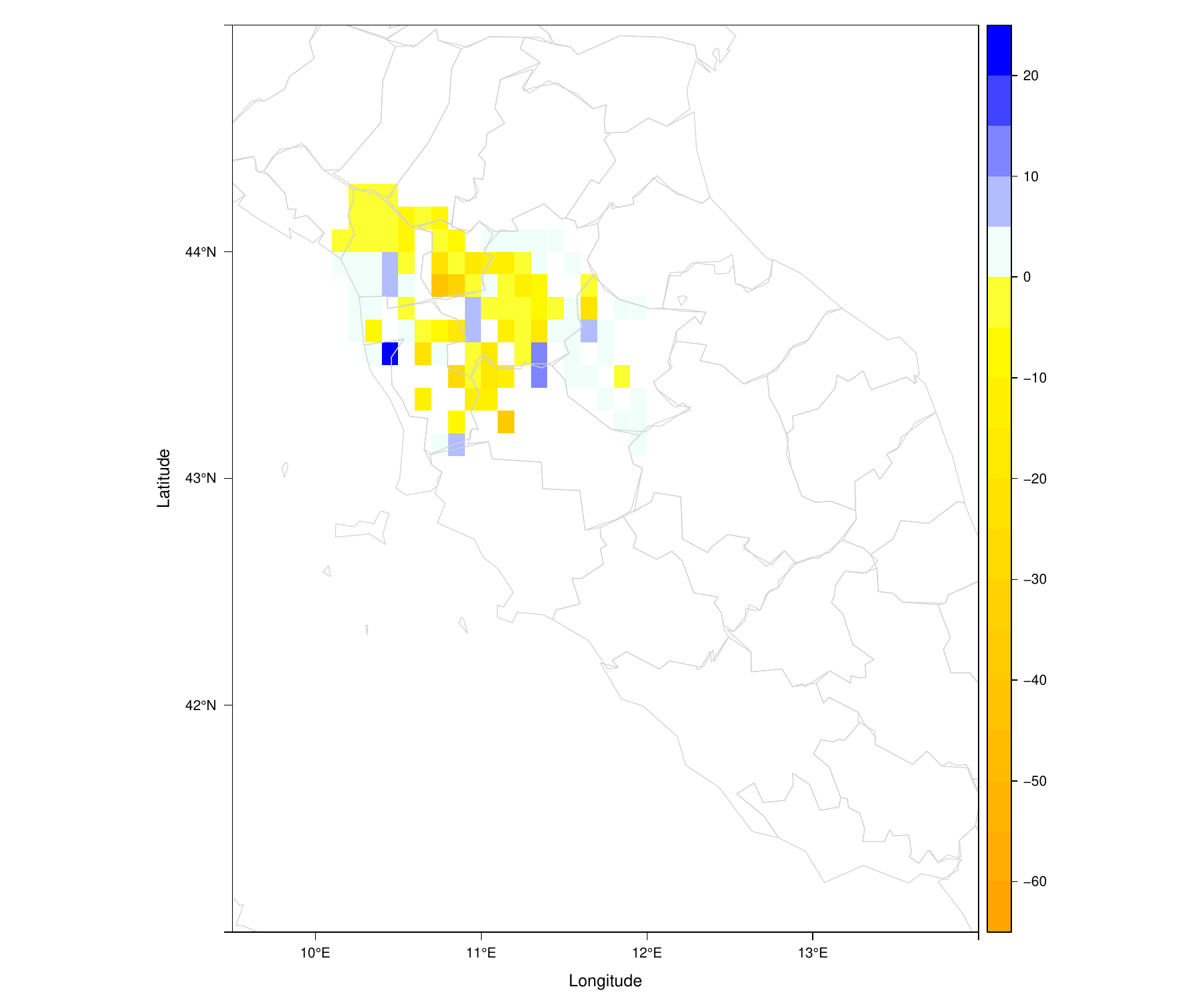}
  } 
  \subfigure[]{
     \includegraphics[width=0.4\textwidth]{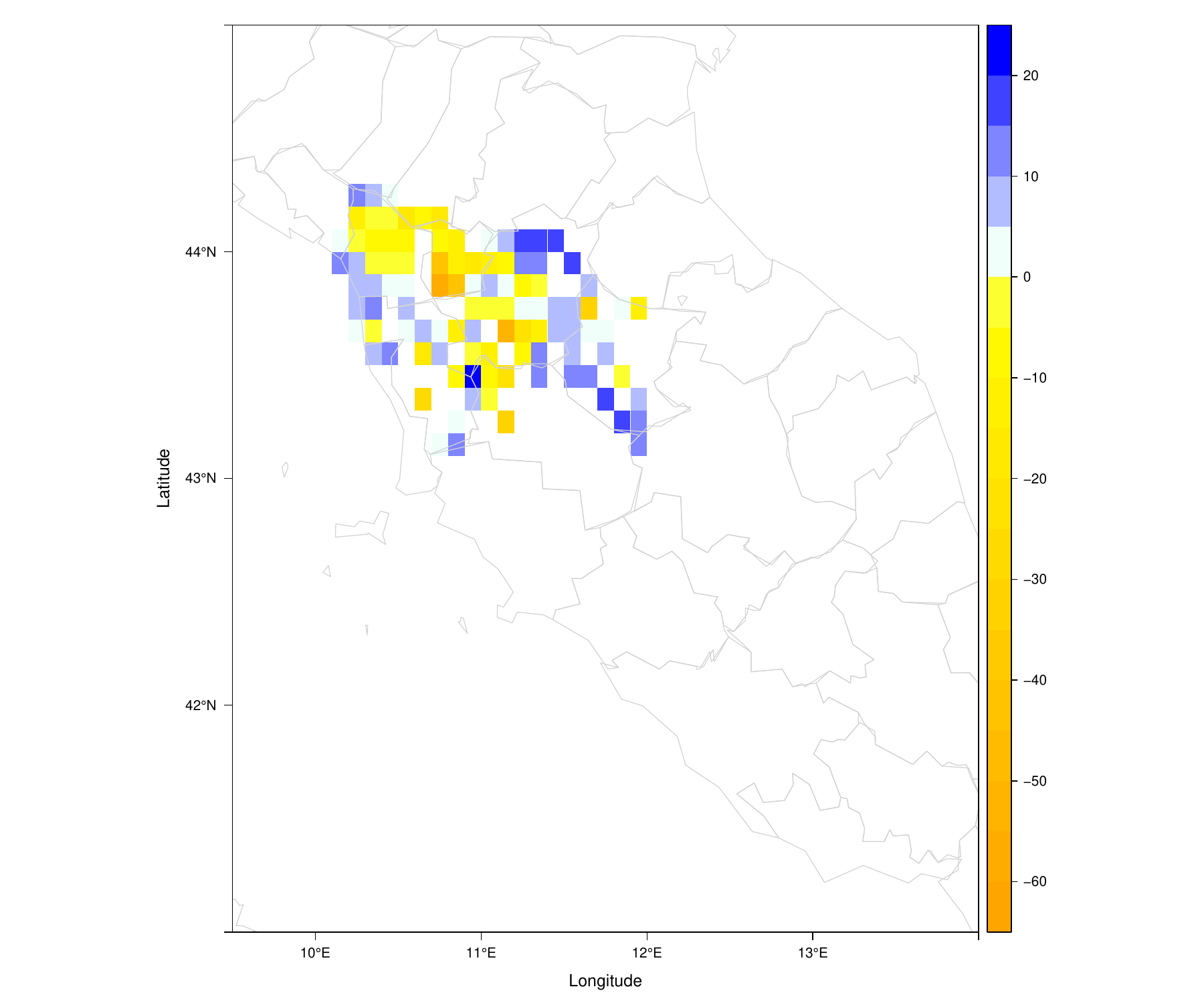}
  }\\
    \subfigure[]{
     \includegraphics[width=0.4\textwidth]{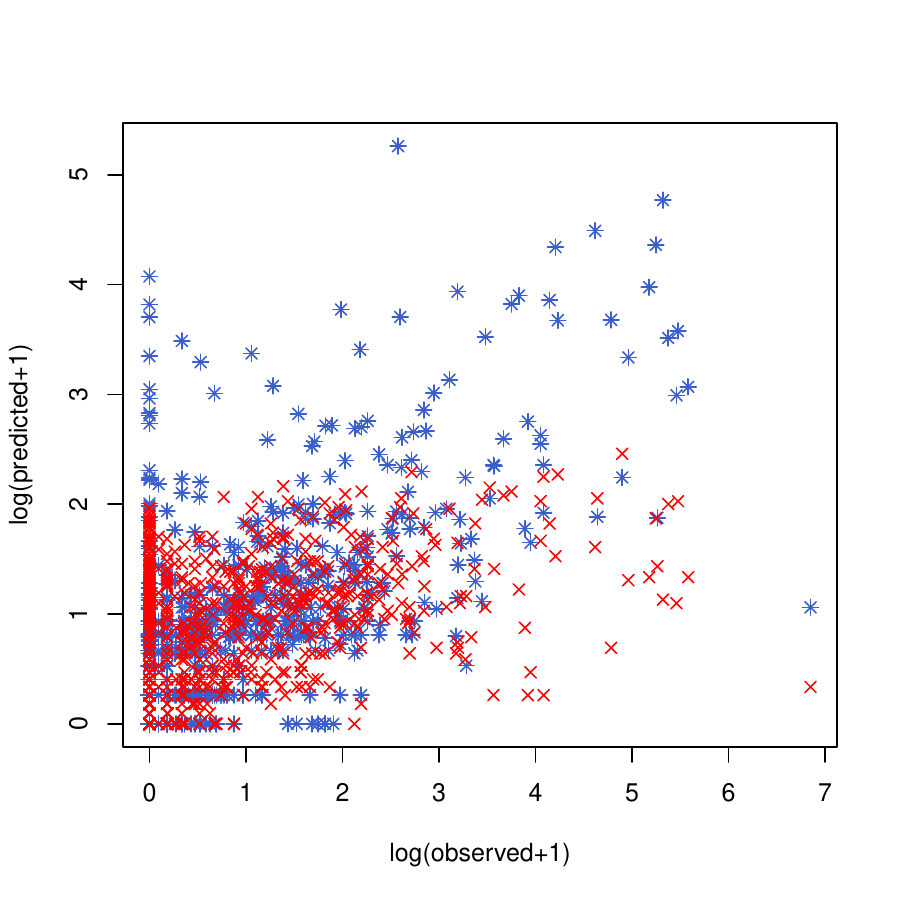}
  }
    \caption{A comparison of M3 predictions and GSMap satellite estimates for the event August 5, 2004 (30-minutes time scale): (a) estimate of errors  probability density function blue for M3 and red for GSMap) with respect to rain gage observations at 1-hour time resolution; (b) and (c) the errors spatial distribution obtained by summing over the entire duration of the event for M3 and GsMap, respectively; (d) log predicted versus log-rain gage observations at 1-hour time resolution (M3 blue, GSMap red). 
    }
    \label{fig:comparison_GSMap_5Aug2004}
  \end{center}
\end{figure}

\begin{figure}
  \begin{center}
  \subfigure[]{
  	\includegraphics[width=0.5\textwidth]{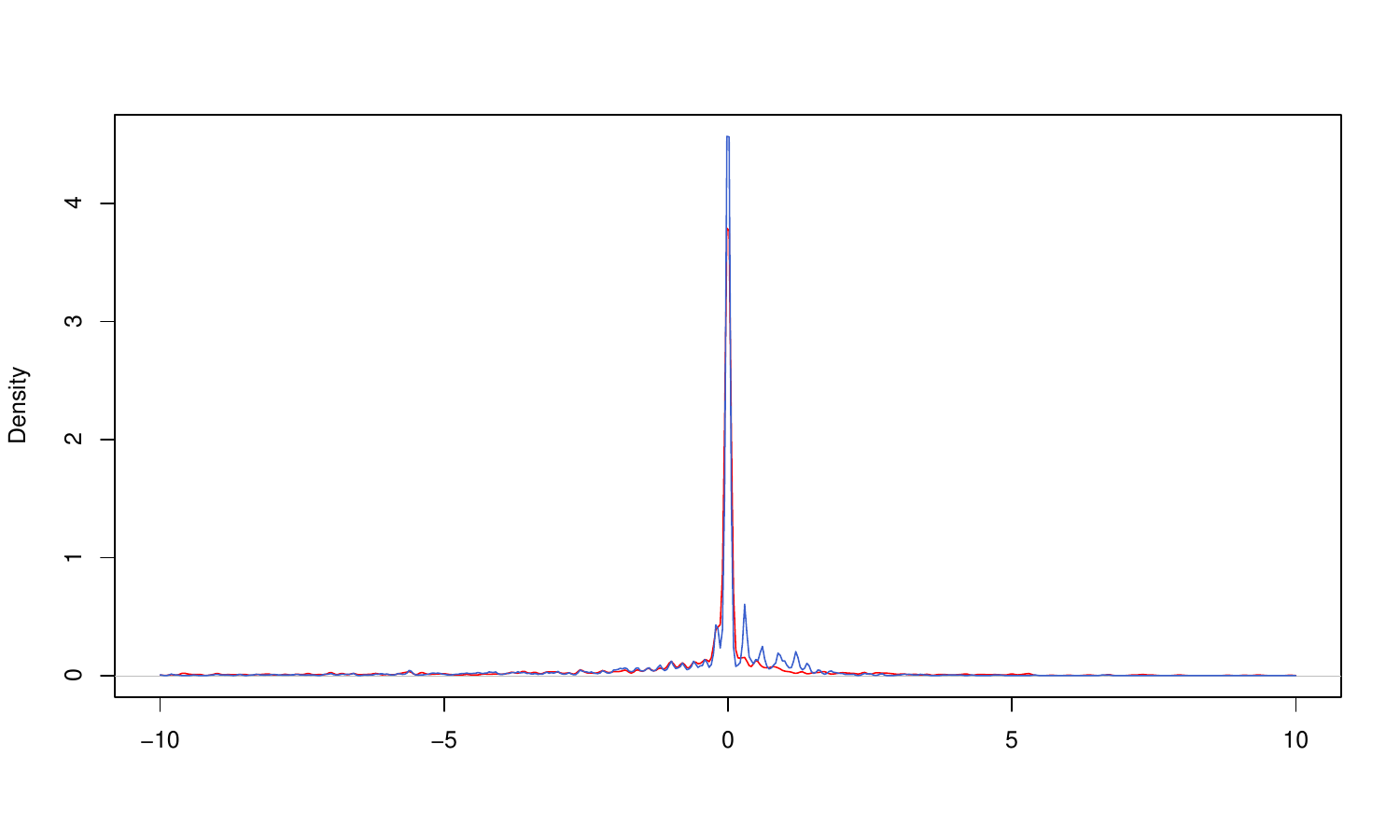}
  }\\
    \subfigure[]{
    \includegraphics[width=0.4\textwidth]{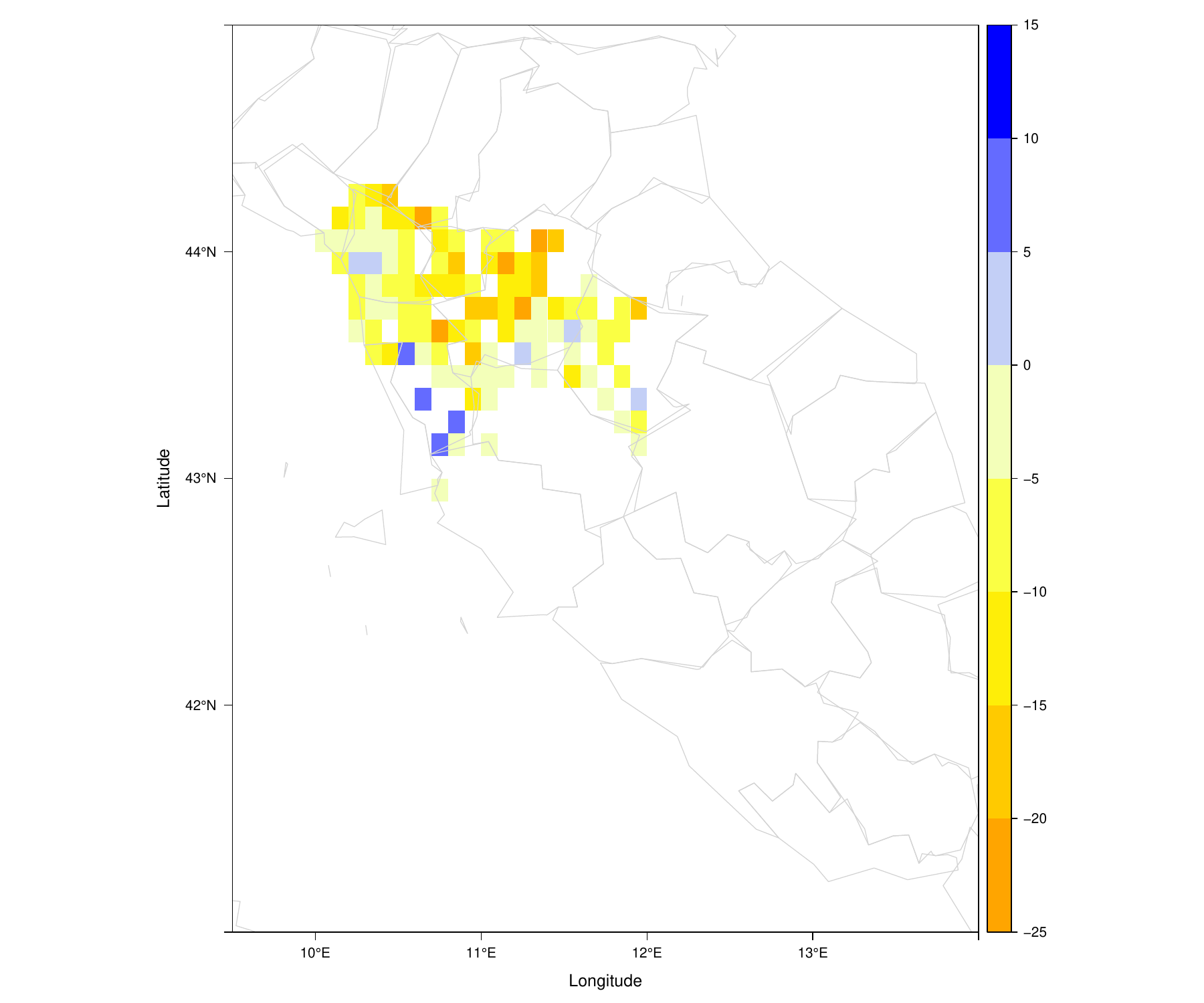}
  }
  \subfigure[]{
     \includegraphics[width=0.4\textwidth]{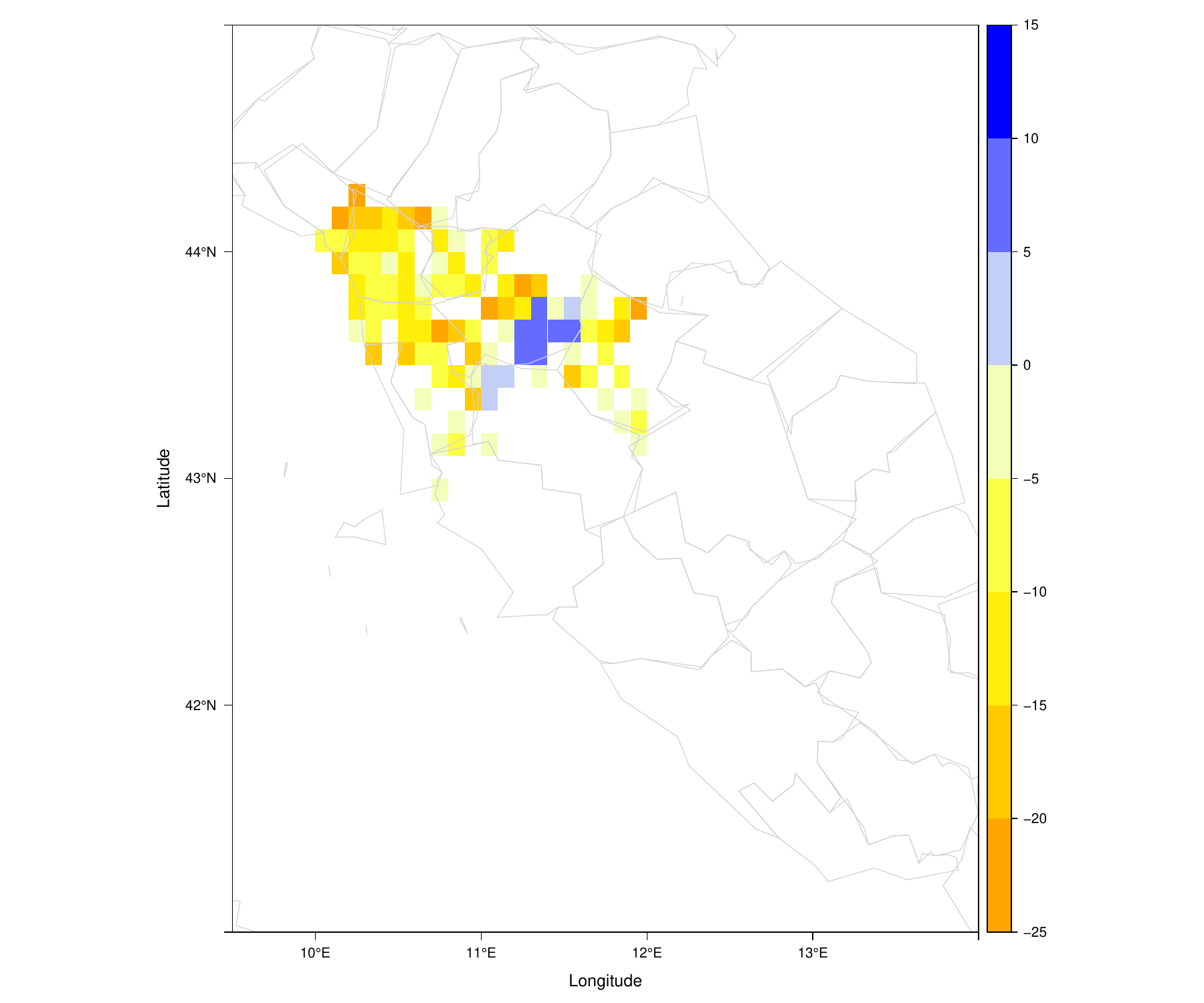}
  }\\
   \subfigure[]{
        \includegraphics[width=0.4\textwidth]{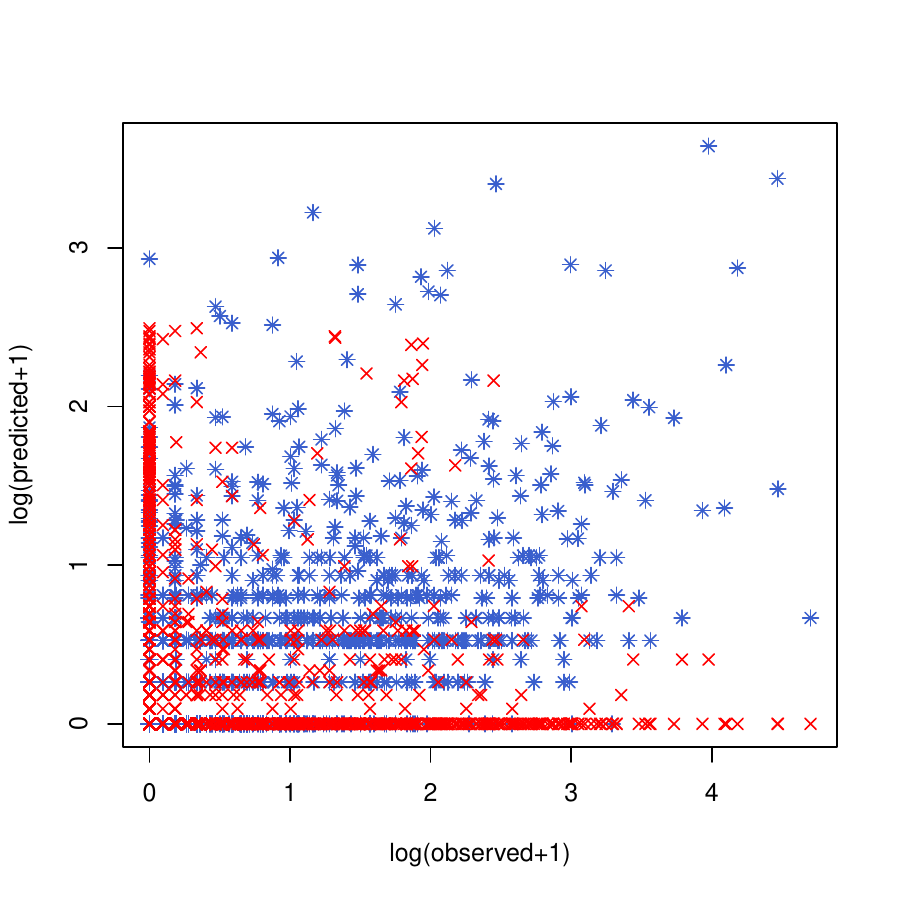}
      }
  \caption{A comparison of M2memory predictions and GSMap satellite estimates for the event May 9, 2006 (15-minutes time scale): (a) estimate of errors the probability density function (blue for M2memory and red for GSMap) with respect to rain gage observations at 1-hour time resolution; ((b) and (c) the errors spatial distribution obtained by summing over the entire duration of the event for M2memory and GSMap, respectively; (d) log predicted versus log  rain gage observations at 1-hour time resolution (M2memory blue, GSMap red)}
  \label{fig:comparison_GSMap_9May2006}
  \end{center}
\end{figure}

We are interested in verifying if any improvement is obtained by our best models predictions on  GSMap satellite estimates \citep{db-satellite}. Since  temporal resolution of the latter is \emph{1-hour}, we aggregate model estimates to the same time scale. We use rain gages observation as benchmark for the comparison taking into account those cells where rain gages are present computing 1-hour cumulated rain.  We compare residuals distributions (the difference between model output and rain gage recordings panel (a) in Figure \ref{fig:comparison_GSMap_5Aug2004} and \ref{fig:comparison_GSMap_9May2006}), the spatial distribution of the difference between estimated events volume and observed one, where volumes are obtained by integrating over time in each cell both predictions and rain gauge recordings and predicted versus observed performance(panels (c)-(d) in Figure \ref{fig:comparison_GSMap_5Aug2004} and \ref{fig:comparison_GSMap_9May2006}),
together with predicted versus observed scatter plots shown in panel (d) of Figure \ref{fig:comparison_GSMap_5Aug2004} and \ref{fig:comparison_GSMap_9May2006}). \\
The median residuals for August event at 30-minutes time scale, are $0.00$ and $0.20$ for M3 and GSMap respectively,  their variances strongly differ as shown in Figure \ref{fig:comparison_GSMap_5Aug2004} (a) (variances are $18.48$ for M3 and $25.22$ for GSMap) suggesting  more accurate predictions returned by M3. As far the May event at 15-minutes time scale  is concerned both models return a zero median residual, but again GSMap produces more variable estimates: variance of M2memory $4.44$, variance of GSMap $7.74$ (see Figure \ref{fig:comparison_GSMap_9May2006} panel (a)). \\
The spatial distribution  (Figure \ref{fig:comparison_GSMap_5Aug2004}(b) and (c)) suggest a better predictive performance of the proposed model than the GSMap, indeed most M3 residuals are closer to zero than those of GSMap.  
Moreover, notice that both M3 and GSMap show a slight overestimation that seems stronger for GSMap. The log-predicted versus log-observed plot highlights how M3 better captures larger values than GSMap while both have a similar behavior in predicting small rain amounts (Figure \ref{fig:comparison_GSMap_5Aug2004} panel (d). Similar considerations can be done when comparing M2memory and GSMap for the May event at 15-minutes, again a better performance of the proposed modeling approach with respect to GSMap (Figure \ref{fig:comparison_GSMap_9May2006} panel (d)).

\section{Discussion and Concluding remarks}
\label{sec:Conclusions}
The main goal of this work is to build a working protocol that allows to isolate rainy events and to  predict the accumulated precipitation at unobserved locations and times  using lightning counting.  The first contribution is a simple and effective scan statistic procedure used for the identification of single storms among several severe meteorological events. This procedure let us to identify $767$ convective events in Central Italy during the period March-September of 2003-2006 from which we estimate the Rainfall Lightning Ratio. As a consequence of the identification process, we are also able to delineate the life-time phases of a convective event 
using the beginning, peak and ending times of lightning temporal evolution to separate them. This aspect represents an advantage of our approach since it let us to incorporate into the model the different features of rainfall propagation in time assuming different weight structures in the equation which converts lightning into rain. Furthermore, as we have a large  number of identified events that we can reliably classify distinguishing among Small, Medium and Large events in terms of number of lightning. This allows us to estimate a different RLR for each event size. 
The validation of the scan statistics procedure is fully carried out  for $7$ events and we believe it requires some further investigation using clouds micro-physical features as a validation tool.

The two convective events of August 5, 2004 and May 9, 2006 chosen for testing our model are both classified in the category of \emph{Large events}. This fact have determined the adoption of the same value of RLR in the estimation of rainfall predictions. Nevertheless, the phenomenological features of the two convective events are substantially different from one another both in lightning intensity and in rainfall rates. In particular, the rain rates and volumes of the August 5, 2004 are larger than those of the  May 9, 2006 event as well as the lightning intensity. The different models predictive performance may depend on the differing events characteristics. 

We estimate our model in $7$ different modifications depending on the type of temporal and spatial weights adopted in the model mean description; three versions are based on lifetime phases of lightning propagation (M1,M2 and M3); two versions are based on single cell activity (M4, M5);  one version rely on both lifetime phases and cell activity (M6) and one version include a different spatial weights specification (memory). Modifications have consequences on model application since the two versions based on cell activity and the memory one can be potentially adapted  in near-real time services when combined as they do not depend on knowledge of the entire storm behavior. The others are \emph{ex-post} estimation models, that can be successfully used in improving historical databases. In particular estimates in midlatitude areas may benefit from our approach.
Predictive intervals have empirical coverage closer to the nominal values of $90\%$ for all the models confirming the accuracy of estimation of our modeling approach. Moreover, it appears that the proposed model is able to capture the time dynamic in all considered situations, remarking that peaks are better predicted when applied at 15-minutes aggregation of event May 9, 2006. In general, our model shows a better performance in fitting the data when applied on August 5, 2004 particularly at 30-minutes aggregation. In terms of RMSE, we can conclude that estimates are more reliable for the more intense event of August 5, 2004 reinforcing the intuition that RLR based estimates work better for intense events. The evaluation done using Probability of Hits on Total, Probability of Detection, Probability of False Detection and HSS confirm  that larger events are better predicted.

On the whole, our modeling approach shows a good capability in capturing time dynamic although it is unable to capture sudden changes in the rainfall series, that is a natural drawback of using a stationary model for time dynamic description.  
Another relevant point is that  rainfall data are typically zero inflated then a model modification, such as using a  mixture between a distribution that specifies the probability of having positive precipitation and a probability density function for the rainfall accumulation may improve the prediction \citep[for instance][]{Berrocal:2008}.   Further developments will include the adoption of a non separable space-time variability structure, that although will add computational complexity it might considerably improve the quality of prediction.\\

\section{acknoledgment} This work has been partially supported by ``AGROSCENARI'' project. The authors wish to thank the ``Consorzio Lamma''  and Luca Fibbi for providing the data and Gianluca Mastrantonio for interesting and useful discussion.

\end{document}